\documentclass[article]{aa}
\usepackage{graphicx}
\usepackage{natbib}
\usepackage{amssymb}
\usepackage{txfonts}
\usepackage{xcolor}
\usepackage{soul}
\usepackage{amsmath}
\usepackage{mathrsfs}
\usepackage{mathtools}
\usepackage{txfonts}
\usepackage{hyperref}
\usepackage{ulem}
\hypersetup{
colorlinks=true,
linkcolor=blue,
citecolor=blue,
filecolor=magenta, 
urlcolor=cyan,
}

\begin{document}

\title{New constraints on the binary system HD~327083 and its gaseous and dusty environments}

\author{L.~Cidale \inst{\ref{1},\ref{2}}\fnmsep\thanks{Member of the Carrera del Investigador Cient\'{\i}fico, CONICET, Argentina} \and
P.~Marchiano \inst{\ref{1},\ref{2}} \and
M.~Kraus \inst{\ref{3}} \and
I.~Andruchow \inst{\ref{1},\ref{2},\star\star} \and
M.~L.~Arias \inst{\ref{1},\ref{2},\star\star} \and
M.~Cur\'e \inst{\ref{4}} \and
M.~Borges Fernandes \inst{\ref{5}} \and
Y.~Aidelman \inst{\ref{1},\ref{2},\star\star} \and
A.~Torres \inst{\ref{1},\ref{2},\star\star} \and
I. Araya \inst{\ref{6}} \and
G.~Maravelias\inst{\ref{7},\ref{8}} \and
J. Zorec\inst{\ref{9},\ref{10}} \and
Y.~Cochetti \inst{\ref{1},\ref{2},\star\star} \and
J. Panei\inst{\ref{1},\ref{2},\star\star}}

\institute{Facultad de Ciencias Astron\'omicas y Geof\'{i}sicas,
  Universidad Nacional de La Plata, Paseo del Bosque S/N, 1900, La Plata, Argentina.\label{1}\\
\email{lydia@fcaglp.unlp.edu.ar}
  \and  
Instituto de Astrof\'{\i}sica La Plata, CCT La Plata, CONICET,
Paseo del Bosque S/N, 1900, La Plata, Argentina.\label{2}
\and
Astronomical Institute, Czech Academy of Sciences, Fri\v{c}ova 298, 251\,65 Ond\v{r}ejov, Czech Republic.\label{3}
\and
Instituto de F\'{\i}sica y Astronom\'{\i}a, Facultad de Ciencias,
Universidad de Valpara\'{\i}so, Av. Gran Breta\~na 1111,
Valpara\'{\i}so, Chile.\label{4}
\and
Observat\'orio Nacional -- MCTI, Rua General Jos\'e Cristino 77,
20921--400 S\~ao Cristov\~ao, Rio de Janeiro, Brazil.\label{5}
\and
Centro Multidisciplinario de F\'{\i}sica, Vicerrector\'{\i}a de Investigaci\'on, Universidad Mayor, 8580745 Santiago, Chile.\label{6}
\and
IASSARS, National Observatory of Athens, Athens, Greece. \label{7}
\and
Institute of Astrophysics, FORTH, Heraklion, Greece.\label{8}
\and
Sorbonne Universit\'e, CNRS, UPMC, UMR7095, Paris, France. \label{9}
\and
Institut d'Astrophysique de Paris, 98bis Bd. Arago, F-75014 Paris, France.\label{10}
}

\date{Received ... / Accepted ...}

\abstract 
   {Binary systems with circumbinary molecular and dust rings are of great interest because they provide insights into the dynamics and evolution of stellar systems and the chemistry of the surrounding material.} 
   {We aim to
   elucidate the nature of the B[e] binary system \object{HD~327083} and constrain its orbital parameters and evolutionary status as well as the physical properties of its circumstellar and circumbinary envelopes.}
   {We present multi-epoch high-spectral resolution optical and near-infrared observations acquired at various observatories combined with VLTI/MIDI observations and ASAS-3 photometry. The data were analysed to obtain the orbital parameters, study individual members of the system, and derive the properties of the atomic and molecular gas and the dusty components
   of the complex environment of \object{HD~327083}.}
 {We improved the orbital solution of \object{HD~327083} and derived a period of $P = 107.699\pm0.005$~d. Most optical lines display variations in shape and intensity when folded with the orbital motion. We assigned F6~II-III and B1 spectral types respectively to the cool and hot stellar components. 
  A ring of molecular gas revolves around the system, and its appearance varies with the orbital phase.
 The MIDI flux distribution shows a silicate band in absorption at $9.7~\mu$m. The interferometric data also indicate that the dust distribution has an elliptical shape, with its inner edge varying with orbital phase between $12.5$~AU and $44$~AU.
 } 
{We conclude that the B[e] binary system \object{HD~327083} comprises a B-type massive accreting star 
and an F-type companion filling over $60\%$ of its Roche lobe. The F-type companion is evolving towards the red supergiant stage.
The deformed shape of the F-type star causes a broad minimum in the light curve. The hot companion ($26\,000$~K, $\log\,g=3.0$) is hidden by a compact disc and shows a bi-polar outflow. The entire binary system presents an O-rich environment. It is surrounded by warm CO and SiO molecular rings, which are enclosed by an elliptical dust structure. All these components were formed by the evolved donor star during its evolution towards the red supergiant phase.}


\titlerunning{The B[e] supergiant HD 327083}
\authorrunning{Cidale et al.}

\maketitle
\section{Introduction}
\vspace{-0.05cm}

Galactic B[e] supergiants (B[e]SGs) are luminous and massive post-main-sequence stars \citep{Lamers1998}. They typically exhibit UV broad blue-shifted absorption lines of highly ionised atoms, narrow permitted and forbidden low-excitation optical emission lines at low expansion velocities, and a strong near/mid-infrared (near/mid-IR) excess. The large infrared emission is attributed to dust particles in the outer parts of the equatorial regions. Some B[e]SGs present high-density equatorial discs that substantially form molecules in a transition zone between the atomic gas and dusty regions \citep{Kraus2019}.

To explain the observed hybrid line spectrum of B[e]SGs, \citet{Zickgraf1985} proposed an empirical model that consists of an evolved B-type star with a hot and fast line-driven wind in the polar region and a much cooler and denser slow wind (of a factor of 10$^2$ to 10$^3$ times higher) in the equatorial zone. Numerical predictions from bi-stable radiation-driven winds in rapidly rotating stars (with $V_{\rm eq}/V_{\rm c} > 0.7$, being $V_{\rm eq}$ the equatorial linear velocity and $V_{\rm c}$ the critical velocity of rotation) support the previous scenario \citep{Cure2005}. These calculations show that a change in the line-force parameters, occurring in the region of the bi-stability jump, might yield density contrasts between polar and equatorial zones of a factor of about 10$^2$ to 10$^4$ in an environment near the stellar surface (r$\,\le$ 2 R$_\star$).

However, evidence increasingly indicates that luminous B[e] stars have detached non-homogeneous circumstellar or circumbinary envelopes in Keplerian rotation \citep{Cidale2012, Marchiano2012, Wheelwright2012a, Wheelwright2012b}, and these discs may present multi-ring structures \citep{Kraus2016, Maravelias2018, Torres2018}. Plausible explanations for the B[e] phenomenon among B-type supergiants are related to either rapid stellar rotation \citep{ Langer1998, Kraus2007, Kraus2008} or mass transfer in binary systems \citep{Podsiadlowski2006, Kraus2013}. In the past few years, several binary systems with primary B[e]SG signatures have been detected (or confirmed) thanks to spectroscopic or interferometric techniques:  \object{V\,921 Sco} \citep{KrausS2012}, \object{MWC\,300} \citep{Wang2012}, and \object{HD 327083} \citep{Miroshnichenko2003}.

We aim to undertake a new study of the binary system \object{HD\,327083} to better constrain its orbital parameters and evolutionary status as well as the physical properties of its circumstellar and circumbinary envelopes (size, density, temperature, and dynamics). A thorough knowledge of binary systems is key to understanding the origin and evolution of stars showing the B[e] phenomenon and the processes leading to molecular and dust disc formation.

This paper is structured in the following way. In Sect.~\ref{star}, we briefly describe our current knowledge of \object{HD\,327083}.  Our observations are described in Sect.~\ref{obs}, and data analysis and modelling are given in Sect.~\ref{results}. In Sect.~\ref{Disc}, we discuss our new results, summarising our main conclusions in Sect.~\ref{conclusions}.
    
\section{The binary}
\label{star}
The object \object{HD\,327083} (J=2000, $\alpha= 17^{h} 15^{m} 15.37^{s}$; $\delta~=~-40^{\circ}~20^{'}~06.79^{''}$, $V = 9.81$~mag, and $K = 3.3$~mag)
is a very reddened P~Cygni-type star with  $E(B-V) \sim 1.9$~mag \citep{Kozok1985}. The star exhibits a strong emission-line spectrum, mainly of \ion{H}{i}, \ion{He}{i} and \ion{Fe}{ii} transitions \citep{Carlson1979}. \citet{Lopes1992} reported that the \ion{He}{i} $\lambda 5875$~\AA\, line was seen in absorption while the \ion{Fe}{ii} and \ion{H}{i} lines displayed P Cygni profiles. 

\citet{Olnon1986}, using IRAS satellite, detected a strong mid-IR flux, with a steep decrease towards longer wavelengths and a featureless spectrum in the $10~\mu$m region. Near-IR spectroscopy also showed strong CO emission features at $2.3~\mu$m \citep{Whitelock1983, McGregor1988} and emission from the SiO first overtone bands arising in the L band \citep{Kraus2015}.

The star's spectral type is still quite uncertain: it was classified into sub-classes B5 \citep{Henize1952},  B8 \citep{Carlson1979}, B6~Ieq \citep{Lopes1992}, and B2 \citep{Miroshnichenko2003}. However, theoretical fittings of the observed H$\alpha$ and H$\beta$ P Cygni line profiles done by \citet{Machado2001} suggested two very different sets of effective temperatures, either $9\,000$ K or $19\,000$ K. Furthermore, both models predicted very high luminosity ($\geq$\,$5\times 10^5$\,$L_{\sun}$) and mass-loss rate ($\geq$\,$4.9\times 10^{-5}$~M$_{\sun}$\, yr$^{-1}$). Later, \citet{Miroshnichenko2003} reported the presence of numerous absorption lines of neutral metals that appear to belong to the photosphere of a moderately cool star. These authors also found RV variations in both emission and absorption lines that vary in anti-phase. They suggested that \object{HD\,327083} should be a binary system seen edge-on, in which the hot (primary) companion was proposed to be a B1-B3 star, while the cool companion (secondary) is an F-type star. For \object{HD\,327083}, they also estimated a distance of about $1.5\pm0.5$~kpc.
\citet{Wheelwright2012a} confirmed \object{HD\,327083} as a Galactic B[e]SG based on the presence of $^{13}$CO emission. Moreover, \citet{Wheelwright2012b} concluded that the CO-forming region originates in a circumbinary Keplerian disc. The inner edge of the best-fitting CO disc is approximately $3\pm0.3$~AU. These authors found no evidence of a variable CO first overtone emission. Based on  VLTI/AMBER observations, \citet{Wheelwright2012a} also supported the binary nature of \object{HD~327083} and the best-fitting model revealed the presence of an elongated circumbinary Keplerian disc where the near-IR excess is originating.
For their analysis, they adopted the distance of $1.5\pm0.5$~kpc derived by \cite{Miroshnichenko2003}. However, updated measurements of the star's parallax provided by Gaia EDR3 place the star at a significantly greater distance, $d=2.448\pm0.145$~kpc \citep{Gaia}. 

 More recently, \citet{Maravelias2018} reported that \object{HD\,327083} displays strong line variations, both in intensity and shape (over the 1999-2016 period) folded with an orbital period of $107.687$~d \citep[see also][]{Nodyarov2024}. In addition, \citet{Maravelias2018}  found that the gaseous component traced by the [\ion{Ca}{ii}] and [\ion{O}{i}] lines lays in the circumbinary rings. The rotation velocities found for the [\ion{Ca}{ii}] and [\ion{O}{i}] forming regions are similar to that of the SiO ring, which implies a common location for these gases. The CO molecule forms (at a slightly higher rotational velocity) in another ring closer to the binary system. 

This binary system was independently and simultaneously studied by two research groups: \citet{Nodyarov2024} and the authors of the present work. This paper also reviews the orbital parameters
and examines the structure and properties of the circumbinary gaseous and dusty discs.

\section{Observations}
\label{obs}

We used different facilities to perform optical and near-IR high-resolution spectroscopic observations of the \object{HD~327083} binary system. Our study was also complemented by analysing mid-infrared interferometric observations and photometric data. All observations were conducted on various epochs between 1999 and 2021.

\subsection{Optical spectroscopic data}

High-resolution ($R\sim 48\,000$) optical observations were obtained using the Fiber-fed Extended Range Optical Spectrograph \citep[FEROS,][]{Kaufer1999}. This spectrograph was attached to the $1.52-$m ESO telescope between  1999 and 2002 and, later, to the $2.2-$m MPG telescope (La Silla, Chile). The science object and sky spectra were simultaneously acquired using the Object-Sky (OBJSKY) mode. All the spectra were reduced with the FEROS pipeline. 

 \begin{table}[h!]
\caption{Log of optical spectroscopic observations of \object{HD~327083}.}
\tabcolsep 0.7pt
\label{table1}
\begin{tabular}{ccccccclcr} 
\hline
\hline \noalign{\smallskip}
  $Date$   &&   $JD$     && $Instr.$ & $Exp.~Time$ &  $Sp.~range$  &&& $\phi$~~ \cr
  [yy-mm-dd] &    & [2450000+] &&           &   [sec]     &  [$\mbox{\AA}$] &&& \cr
\noalign{\smallskip} \hline \noalign{\smallskip}
1999-06-25 && $1354.751$ && FEROS &$1800$& $3585-9180$ &&& $0.45$  \\
2000-04-20 && $1654.750$  && FEROS & $1800$ & $3850-8900$ &&& $0.24$\\
2000-04-22 && $1656.759$ && FEROS & $1800$ & $3850-8900$ &&& $0.26$ \\
2010-05-21 && $5337.779$  && REOSC & $3600$ & $4075-6900$ &&& $0.43$ \\
2010-05-23 && $5339.840$  && REOSC & $3600$ & $4075-6900$ &&& $0.45$\\
2012-04-12 && $6030.357$  &&  B\&C & $1200$ & $3500-5000$ &&& $0.87$\\
2013-06-12 && $6455.718$  && REOSC	& $1800$ & $4900-6900$ &&& $0.81$\\
2014-02-10 && $6698.839$  && REOSC & $1800$ & $4900-6700$ &&& $0.07$ \\
2014-03-22 && $6738.878$  && REOSC & $1800$ & $4900-6700$ &&& $0.44$ \\
2014-04-11 && $6759.870$  && REOSC & $2400$ & $4500-6700$ &&& $0.64$\\
2014-04-14 && $6762.724$  && REOSC & $2400$ & $4500-6700$ &&& $0.67$\\
2015-05-05 && $7147.775$  && REOSC & $2400$ & $4500-6700$ &&& $0.25$ \\
2015-10-12 && $7307.518$  && FEROS &  $500$ & $3580-9150$ &&& $0.72$\\
2015-10-12 && $7307.525$ && FEROS &  $500$ & $3580-9150$ &&& $0.72$ \\
2015-10-15 && $7311.492$  && FEROS &  $500$ & $3580-9150$ &&& $0.76$\\
2015-10-15 && $7311.498$  && FEROS &  $500$ & $3580-9150$ &&& $0.77$\\
2016-04-13 && $7491.660$  && FEROS &  $500$ & $3775-9200$ &&& $0.43$\\
2016-07-28 && $7597.598$  && FEROS &  $600$ & $4000-9000$ &&& $0.41$ \\
2017-06-11 && $7916.003$  && B\&C & 700 & $3500-5000$ &&& $0.37$\\
2017-08-26 && $7991.598$  && FEROS &  $600$ & $4000-9000$ &&& $0.08$ \\
2017-08-29 && $7994.600$  && FEROS &  $600$ & $4000-9000$ &&& $0.10$ \\
2017-09-02 && $7998.525$  && FEROS &  $600$ & $4000-9000$ &&& $0.14$ \\
\noalign{\smallskip} \hline \noalign{\smallskip}
\end{tabular}
\label{table:1}
\end{table}

An additional set of observations was taken with the REOSC echelle Cassegrain spectrograph in cross dispersion mode mounted at the $2.15-$m  Jorge Sahade (JS) telescope at the Complejo Astronómico El Leoncito (CASLEO, San Juan, Argentina). We selected the following instrumental configuration: a $400~\ell$mm$^{-1}$ grating (blazed at $4000$~\AA), a slit width of $250~\mu$m, and a TEK~$1024\times1024$ CCD. Th-Ar comparison lamp spectra were taken to perform the wavelength calibration. The covered spectral region ranges from $4075$~\AA\, to $6900$~\AA\, with a resolving power of $R~=~12\,500$. A standard spectral reduction procedure was applied using IRAF\footnote{IRAF is distributed by the National Optical Astronomy Observatory, which is operated by the Association of Universities for Research in Astronomy (AURA) under a cooperative agreement with the National Science Foundation.} tasks.

Low-resolution spectra were also obtained with the Boller \& Chivens  (B\&C) spectrographs attached to the $1.6-$m Perkin-Elmer telescope at the Laborat\'orio Nacional de Astrof\'isica (LNA), Braz\'opolis, Brazil, on 2012, April 12 and to the JS telescope (CASLEO,  Argentina) on 2017, June 11. For the former spectrum, a grating of $600$~$\ell$mm$^{-1}$, a slit width of $400~\mu$m, and a Marconi CCD~42-40-1-368 ($2048~\times~2048$ pixels) detector were selected. This instrumental configuration provides a $2.12$ \AA\ dispersion every two pixels ($R=1\,200$). For the B\&C spectrum of 2017 ($R \sim 700$), we used the TEK~$1024\times1024$ CCD, a grating of $600$~$\ell$mm$^{-1}$, and a slit width of $350~\mu$m.  Both spectra cover the interval $3500-5000$~\AA. A comparison lamp of He-Ar was used to perform the wavelength calibration. The flux calibration was performed by observing flux standard stars selected from \citet{Hamuy1994}. The reduction procedure was done with IRAF tasks. The `apscatter' task was also used to subtract the background (scattered light and sky).

The log of spectroscopic observations is listed in Table \ref{table:1}: columns 1 and 2 provide observation and Julian dates, while columns 3, 4, 5, and 6 give, in turn, the instrument used, exposure time, wavelength range, and orbital phase (calculated with the orbital period of 107.699~days and the time of conjunction ($T_c$) found in this work, see Sect.~\ref{orbit} and Table~\ref{table-orbit}).

\begin{table}[h!]
\caption{Log of near-IR spectroscopic observations.}
\tabcolsep 1.7pt
\begin{tabular}{lrccccc} 
\hline
\hline \noalign{\smallskip}
~~~~$Date$ &   $JD$~~~~~~~~         & $Instr.$  &  $N \times T_{\rm exp}$ &  $Sp.~range$  && $\phi$   \cr
 [yy-mm-dd]      & [2450000+] &   & [sec]    &   [$\mu$m]  & & \cr
 \noalign{\smallskip} \hline \noalign{\smallskip}
 2010-06-28 &  $5375.704$  & CRIRES  & $20\times 20$ & $2.277-2.325$ && $0.78$\\ 
 2010-08-01 &  $5410.469$  & Phoenix & $4 \times 42$ & $2.319-2.329$ && $0.12$\\
 2017-04-12 &  $7855.845$  & Phoenix & $4 \times 15$ & $2.318-2.328$ && $0.81$\\
 2017-04-12 &  $7855.873$  & Phoenix & $5 \times 15$ & $1.970-2.420$ && $0.81$\\ 
  2021-08-18 &  $9443.481$  & IGRINS  & $4 \times 15$ & $ 1.490-1.780$ && $0.57$\\
2021-08-18 &  $9443.481$  & IGRINS  & $4 \times 15$ & $2.265-2.450$ && $0.57$ \\
\noalign{\smallskip} \hline \noalign{\smallskip}
\end{tabular}
\label{table:2}
\end{table}

\subsection{Light curve}

The light curve of \object{HD\,327083} is available in the All Sky Automated Survey (ASAS) Photometric V-band Catalogue \citep{Pojmanski1997}. Particularly,  ASAS-3 provides light curves for objects south of declination $+28^\circ$ during the years 2000-2009. The light curve has been presented by \citet{Maravelias2018}. Here, it is used to complement the information obtained from spectroscopy.

\subsection{Near-IR spectroscopic observations}

To study the CO molecular band heads around $2.3~\mu$m, we used high-resolution IR spectra ($R\sim 50\,000$) of \object{HD\,327083}, acquired with the visitor instruments Phoenix \citep[Phoenix Infrared High-Resolution Spectrograph,][]{Hinkle2003, Hinkle1998} and IGRINS \citep[Immersion GRating INfrared Spectrometer,][]{Park2014} spectrographs mounted on the $8.1-$m Gemini South telescope (Cerro Pach\'on, Chile). The Phoenix observations were taken in service mode under programs GS-2010A-Q-41 and GS-2017A-Q-30, while the IGRINS spectra were acquired under program GS-2021A-Q-401.  In 2010, Phoenix observations were taken with the $K4308$ filter centred at $2.323~\mu{\rm m}$ and the Aladdin $512 \times 1\,024$ InSb array as the detector. In 2017, the first CO band head was observed with the filters $K4396$ and $K4308$, centred at $2.294~\mu{\rm m}$ and $2.323~\mu{\rm m}$, respectively. The completely covered wavelength intervals are $2.289-2.299~\mu$m and $2.319-2.328~\mu$m. The spectrograph IGRINS  provides a resolving power of $R \sim 45\,000$ with a wavelength range in the H and K bands of $1.49-1.78~\mu$m and $1.97-2.42~\mu$m, respectively.

An offset pattern ABBA (science-sky-sky-science) nodding along a $4-$pixel wide slit was applied to remove the sky emission. 
Phoenix and IGRINS spectra were reduced using IRAF software routines and the IGRINS pipeline package\footnote{https://github.com/igrins/plp}, respectively. The basic reduction steps were subtraction of the AB pairs, flat fielding, wavelength calibration, and telluric correction. To apply the telluric correction, we selected a B late-type (B9~V) standard star observed immediately before (or after) our science object. In all the cases, the telluric features were removed with the {\it telluric} task from the IRAF packages. The final spectra were normalised to the continuum.

To supplement the data analysis, we also used the spectra available in the whole wavelength range $2.277-2.325~\mu{\rm m}$ taken with the CRyogenic high-resolution InfraRed Echelle Spectrograph \citep[CRIRES;][]{Kaeufl2004} mounted on the $8.2-$m telescope at ESO-VLT (Paranal, Chile). The reduction procedure is described in \citet{Maravelias2018}, who discussed only the first CO band-head spectrum. 

Table \ref{table:2} gives the log of the near-IR spectroscopic observations: columns 1 and 2 provide the observation and Julian dates, respectively. Columns 3, 4, 5, and 6 give information on the instrument used, the exposure time, the covered wavelength range, and the phase related to the orbital motion.

\begin{table}[h!]
\begin{center}
\caption{Observation log of VLTI/MIDI.}
\tabcolsep 3.8 pt
\begin{tabular}{cccccc}
\hline\hline
\noalign{\smallskip}
 $Date$  &  $T_{\rm obs}$ &  $UT$  &  $B_{\rm p}$  &  $PA$ & $\phi$\\
yy-mm-dd &  [UTC] &  baseline &  [m] &  [$^{\circ}$]&\\
\noalign{\smallskip} \hline \noalign{\smallskip}
2010-04-25 & 09:46:09 & $UT3-UT4$ & $60.5$ & $131.1$ & $0.19$\\  
2010-04-26 & 09:35:06 & $UT2-UT3$ & $41.1$ & $54.6$ &$0.20$\\   
2010-05-25 & 03:59:36 & $UT2-UT3$ & $46.6$ & $25.6$ & $0.47$\\   
\noalign{\smallskip} \hline \noalign{\smallskip}
\end{tabular}
\label{table:3}
\end{center}
\end{table}
\vspace{-0.6cm}
\subsection{Interferometric observations}

 The object \object{HD\,327083} was observed at the Paranal ESO Observatory (Chile) in service mode (ESO Program ID: 085.D-0454(B)) with the Very Large Telescope Interferometer (VLTI) and the MID-infrared Interferometric instrument \citep[MIDI;][]{Leinert2004}. MIDI is a Michelson-type interferometer with a half-reflecting plate optical recombiner. It combines signals from two telescopes of the Very Large Telescope (VLT) array, using either two $8.2-$m unit telescopes (UTs) or two $1.8-$m auxiliary telescopes.

Three datasets were taken at different projected baseline lengths ($B_p$) and position angles ($PA$) in 2010 April, and May, using the $UT2-UT3$ and $UT3-UT4$ baselines. Table \ref{table:3} summarises the observation log: date, time of observation (Coordinated Universal Time = UTC), the configuration of the interferometer, projected baseline lengths ($B_{\rm p}$), position angles of the projected baseline on the sky ($PA$), and the orbital phase ($\phi$) of the observation date calculated using the period of the binary system obtained in this work.

Both $N$-band spectrum and the spectrally dispersed fringes were recorded between $8.4\,\mu$m and $13\,\mu$m with a spectral power of $R=30$. Data reduction was done with MIA+EWS packages \citep{Jaffe2004, Leinert2004}. As calibrator stars, we used  \object{HD\,159433}: $\theta_{\mathrm{UD}}~=~2.07~\pm~0.13$~mas, and $F_{\mathrm{12}} = 9.27$~Jy and \object{HD\,152980}: $\theta_{\mathrm{UD}}~=~3.64~\pm~0.12$~mas, and $F_{\mathrm{12}} = 22.03$~Jy\footnote{The angular diameter, $\theta_{\mathrm{UD}}$, was obtained from the calibrators list provided by MIA+EWS data reduction package and the flux, $F_{\mathrm{12}}$, at the [12] band was taken from \citet{Cidale2012}.}.

\section{Results}
\label{results}

\subsection{Spectral classification and stellar parameters}
\label{sect:BCD}

\begin{figure*}[h]
\centering
\includegraphics[width=0.9\hsize,angle=0]{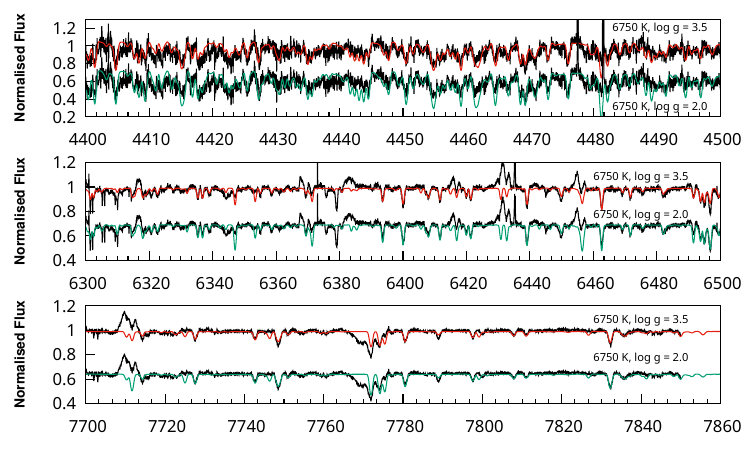}
\qquad
\caption{Spectrum from FEROS of \object{HD~327083} taken on 2017, August 26 (solid black line) compared with MARCS atmospheric models of $6600-6750$~K for $\log\,g=2.0$ and $\log\,g=3.5$ (solid green and red lines, respectively). 
This region is dominated by photospheric lines (of \ion{H}{i} and metals) from a late F-type star. The emission lines are formed in the disc of the hot companion:  \ion{He}{i} (P Cygni profiles), H$\alpha$, and \ion{Fe}{ii} lines.} 
\label{optical_feros}
\end{figure*}

\begin{figure*}[h]
\begin{center}
\includegraphics[width=12cm,angle=270]
{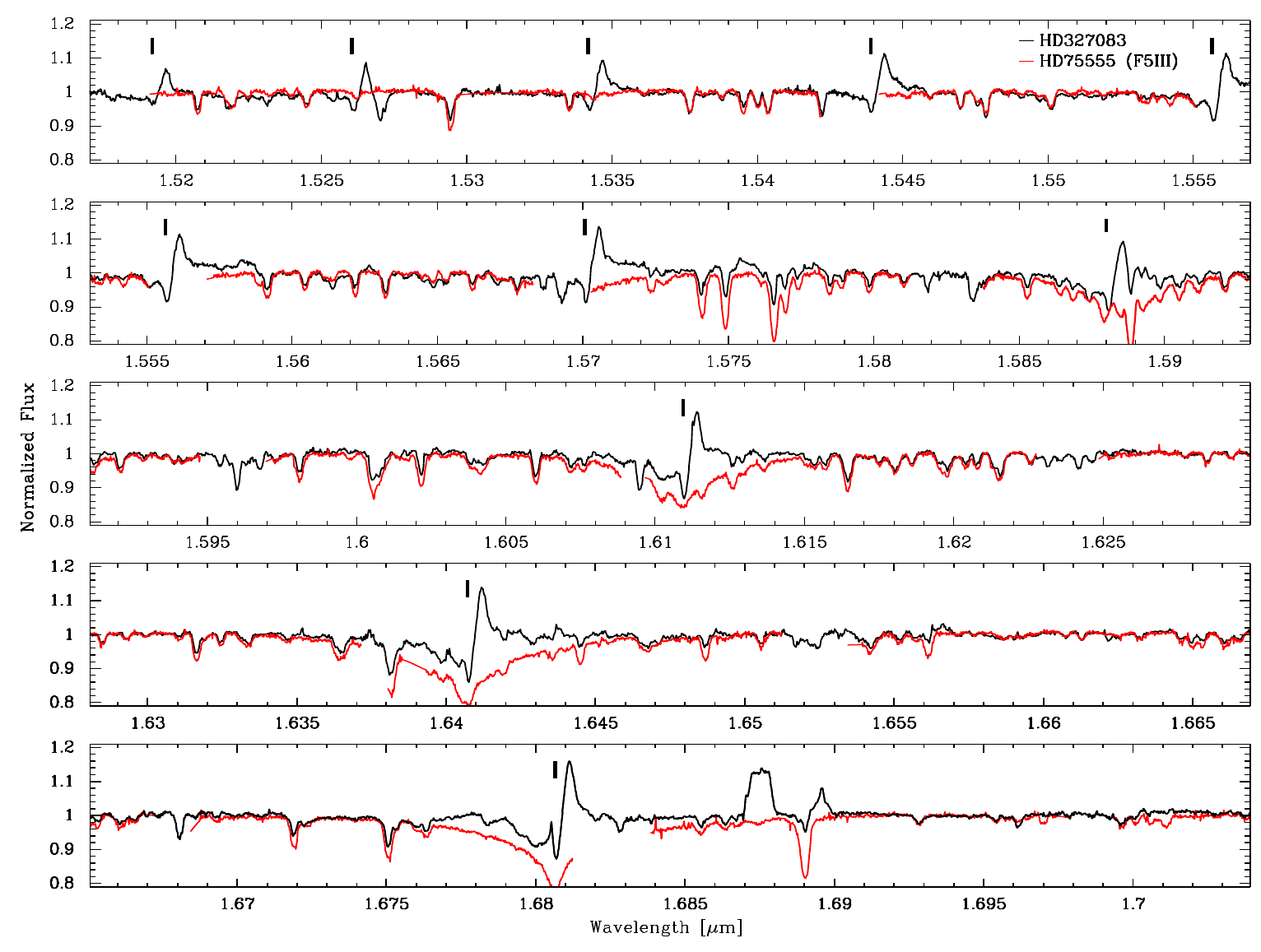}
 \caption{Portion of the IGRINS H-band spectrum. This region is dominated by photospheric lines (of \ion{H}{i} and metals) from an F-type star. Emission lines with P~Cygni profiles correspond to \ion{H}{i} (indicated with ticks) and \ion{Fe}{ii} (seen in the 1.687-1.690 $\mu$m region). These lines are formed in the disc around the B-type companion. For comparison purposes, an F5~III star is overplotted (shown in red).} 
 \label{fig-H-band}
 \end{center}
\end{figure*}

The FEROS optical spectra present numerous metallic absorption lines of neutral and singly ionised elements (e.g. \ion{Si}{ii}, \ion{Ca}{i}, and \ion{Fe}{i}) typically observed in a mid-F-type star. All the absorption lines show RV variations. 
Superimposed on the F-type spectrum are prominent double-peaked emissions of \ion{Fe}{ii} lines and P~Cygni profiles from \ion{H}{i} and \ion{He}{i}, indicating the presence of a gaseous disc around a hot companion and a (most likely) bipolar wind seen under some inclination angle relative to the line of sight.  We could not detect any purely photospheric absorption line corresponding to the hot star.

The FEROS spectra were compared with solar-abundance synthetic spectra generated using MARCS model atmospheres and the TURBOSPECTRUM synthesis code, as part of the AMBRE project \citep{Laverny2012}. These synthetic spectra are accessible through the POLLUX database\footnote{http://pollux.oreme.org} \citep{Palacios2010}. 
The best fit to the normalised optical spectrum of
\object{HD~327083}, near the minimum brightness (see Sect.~\ref{ASAS} for the light curve analysis), was achieved for a $T_{\rm eff}=6750$~K and $\log\,g=3.5$ (see Fig.~\ref{optical_feros}, red curves compared to observations in black). An atmospheric plane-parallel model with a lower surface gravity ($\log\,g = 2.0$, green solid line) predicts deeper absorption lines. Models with lower values of $\log\,g$ are discarded, as they provide poor fits to the observed spectra.
 The discrepancy between each model and the observation is subtle. A model with $\log\,g=2.0$ tends to better fit the spectral lines that are more sensitive to luminosity. This model is also more consistent with the evolutionary state inferred for the binary system (see Sect~\ref{nature}). However, none of the models match the intensity of the \ion{O}{i} triplet, which shows contamination with a broad feature.  All synthetic spectra were modified to match the FEROS spectral resolution and line broadening to account for the star's projected rotational velocity, requiring a value of  $v \sin~i = 22$~km~s$^{-1}$.

Photospheric absorption lines of an F5 or F6 III-type star are also noticeable in the IGRINS spectrum taken in the near-IR region. Figure~\ref{fig-H-band} illustrates the H-band spectral region of \object{HD\,327083}, which is compared with an F5~III star taken from the IGRINS library. In this spectral region, we also observed a combined spectrum of photospheric absorption Pfund lines from the cool star and emissions of \ion{H}{i} arising from a gaseous disc. The \ion{He}{i} lines have P~Cygni profiles.   

\begin{figure}[h]
\hspace{-0.4cm}\includegraphics[width=9.2cm,angle=0]{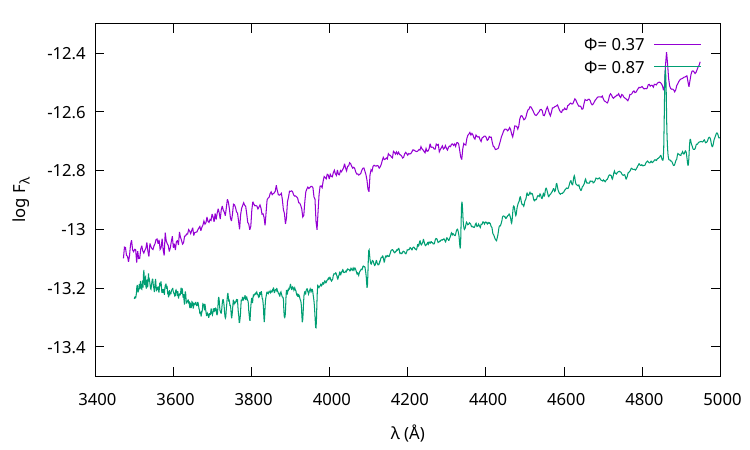}
 \caption{Low-resolution optical spectrum at different phases showing variation in the line intensity and the Balmer continuum.}
 \label{BCD}
\end{figure}

The B\&C low-resolution spectra are shown in Fig.~\ref{BCD}\footnote{The flux $F_\lambda$ if given in erg~cm$^{-2}$~s$^{-1}$~\AA$^{-1}$}. The appearance of the spectrum changes with orbital motion. The spectrum taken in 2012, at $\phi=0.87$, shows a hot source (revealed through a negative Balmer continuum slope) and \ion{H}{i} lines with P~Cygni profiles. The spectrum from 2017 (taken at $\phi=0.37$) does not show a negative Balmer continuum slope nor prominent P~Cygni line profiles. Only H$\beta$ and some \ion{Fe}{ii} lines are observed in emission. The spectra present a small (or even absent) Balmer jump, $D \sim 0.03$ dex \citep[D is the height of the Balmer discontinuity according to the BCD spectrophotometric classification system,][]{Chalonge1973, Zorec2009, Zorec2023} which is consistent with late-F or early B-type stars. The presence of photospheric \ion{Ca}{ii} lines (being the K component blended with a Balmer line) also suggests a mid-F type star. These characteristics reveal that the dominant continuum radiation source comes from a cool star. 

\begin{figure}[h]
\begin{center}
\includegraphics[width=6cm,angle=270]{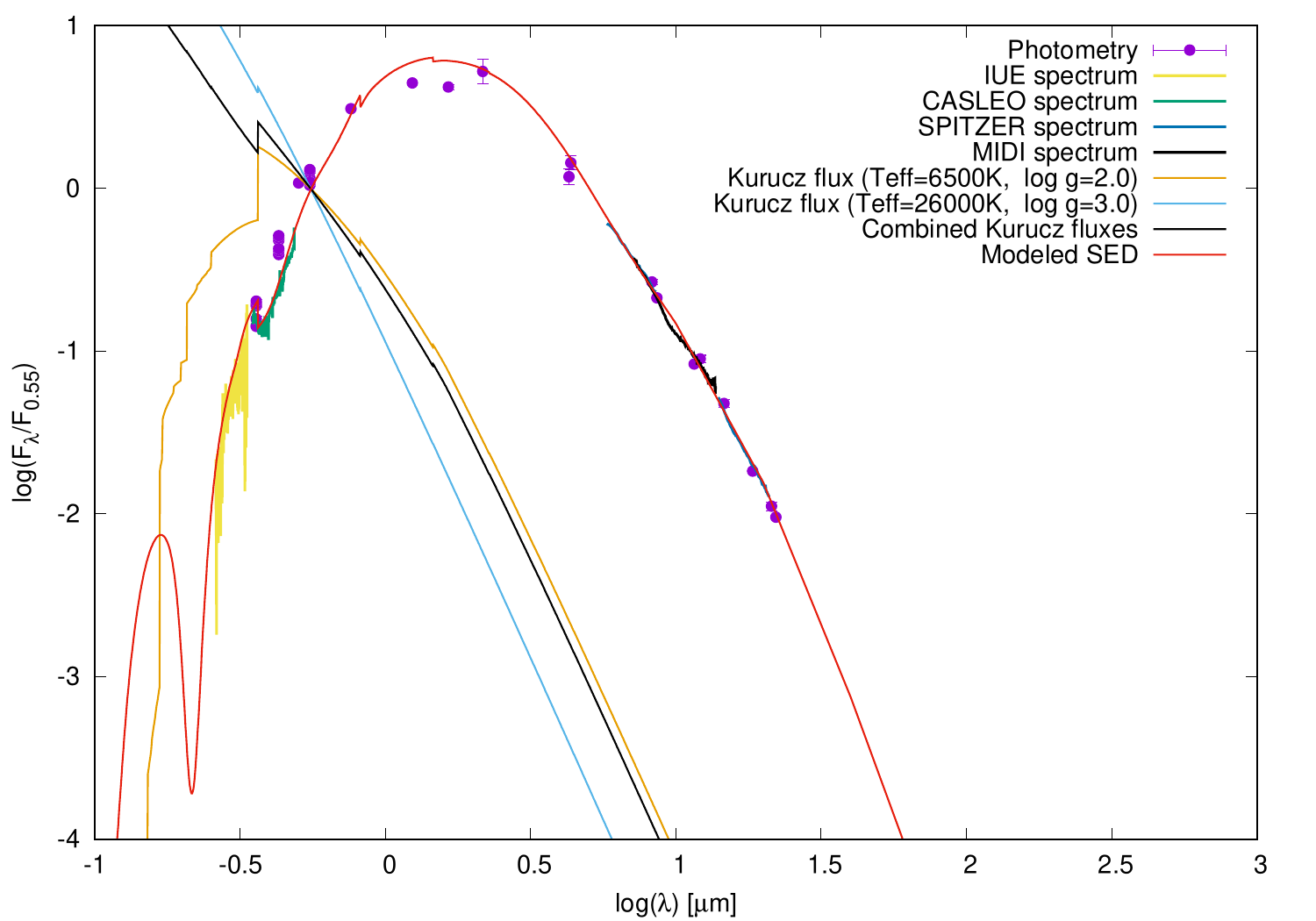}
 \caption{Spectral energy distribution and the best-fitting model (in solid red line, see details in the text).  Photometric data were taken from \citet{Kosok1985, Cutri2003, Egan2003, Miroshnichenko2003,  Ishihara2010, Cutri2012, Gaia, Marton2024}. The Spitzer and IUE spectra were taken from their corresponding archives.} 
 \label{Fig_SED}
 \end{center}
\end{figure}

Figure~\ref{Fig_SED} shows the spectral energy distribution (SED) of \object{HD~327083}, which is composed of different photometric bands, our B\&C and MIDI spectra, together with observations from the Spitzer Heritage Archive and the International Ultraviolet Explorer (IUE). 
To model the SED, we follow the works of \citet{Zorec1998},  \citet{Muratore2011},  and \citet{Marchiano2012}, where the stellar object is surrounded by a gaseous shell, characterised by a mean or effective radius $R_{\rm G}$, near the star, while dust shells, of radii $R_{\rm D}$, lie farther out. Each shell contributes to the attenuation of stellar radiation and re-emits energy according to its local physical conditions. Assuming that the geometrical depth of each shell is smaller than its distance to the star, $R_{\rm G}$ or $R_{\rm D}$, the emergent intensity can be calculated by applying a plane-parallel radiative transfer solution \citep[see details on the geometrical model in][]{Cidale1989, Moujtahid1999, Marchiano2012}. To model the central stellar object, we combine the spectral energy distribution of the underlying cool and hot stars by scaling fluxes from the atmospheric models of \citet{Castelli2003},  assuming various radius ratios. 

Since our SED calculation is based on a simplified model, it does not include the variations in the extinction profile expected for different dust components. Thus, the optical depths of gas and dust are calculated using different total-to-selective extinction ratios ($R_{\rm V}= A_{\rm V}/E(B-V)$), and the extinction law given by \citet{Cardelli1989}. Dust shells are assumed to be optically thin, and the temperature distribution is described using a power-law relation, $T_{\rm D}(r)= T_{\rm eff}\,(2\,r/R_\star)^{-2/(4+p)}$. The parameter $p$ accounts for the absorption properties of dust grains \citep[typically $p=1$, see][Eq.~2.27]{Lamers1999}. 
 
The best fit to the SED was achieved for atmospheric models of a cool star of $6\,750$~K along with a hot star of $26\,000$~K and a hot-to-cool component radius ratio of  $R_2/R_1=0.14$.  We adopted $\log\,g = 2.0$ and $\log\,g=3.5$ for each star, respectively. We also derived a total colour excess of $E(B-V)=1.4$~mag, assuming $R_{\rm V}=3.1$, which is consistent with a diffuse interstellar medium. A gaseous shell is found at $R_{\rm G}=1.1$~R$_1$ (where
R$_1$ is the radius of the cool star) with a $T_{\rm G}=16\,000$~K. This gaseous shell produces an attenuation of $0.06$~mag. Its main effect is to enhance the continuum flux in the spectral region where the maximum occurs. 
Four dust shells of radius $R_{\rm D}$  were necessary to fit the mid-infrared region accurately ($R_{\rm D}= 30$~R$_1$, $T_{\rm D}=1\,360$~K; $R_{\rm D}= 60$~R$_1$, $T_{\rm D}=1\,030$~K; $R_{\rm D}= 150$~R$_1$, $T_{\rm D}=715$~K; $R_{\rm D}= 1\,700$~R$_1$, $T_{\rm D}=270$~K). To model the dust component, we required a value of $R_{\rm V}=3.7$. 
 Therefore, dust grains appear to be of circumbinary origin and larger than the typically interstellar ones, resulting in an additional extinction of $\sim 0.9$ mag.

In summary, based on the optical and IR data study, we assigned the spectral types F6~II-III ($T_{\rm eff} = 6750\pm 150$~K, $\log\,g=2.0\pm0.5$) and B1 ($T_{\rm eff} = 26\,000\pm 2500$~K, $\log\,g=3.0\pm0.5$) to the binary system's cool and hot star components, respectively. The binary system is surrounded by dust composed of processed material. The luminosity and masses of the binary components are discussed in Sect.~\ref{nature}.

\subsection{Orbital parameters}
\label{orbit}

From the analysis of FEROS and CASLEO spectra, we derived average values of radial velocity (RV) measurements corresponding to the photospheric metallic lines of \ion{Si}{ii}~$\lambda\lambda~6347, 6371$~\AA, \ion{Ca}{i}~$\lambda~6162$~\AA, \ion{Ca}{i}~$\lambda\lambda~6439,6449,6462,6471$~\AA, and \ion{Ca}{i}~$\lambda~6717$~\AA, which are given in Table \ref{table-RV}, column 4. The RVs of the emission line \ion{Fe}{ii}~$\lambda6084$~\AA\, are listed in column 5. 

\begin{table}[h!]
\begin{center}
\caption{Radial velocity measurements at different orbital phases.}
\tabcolsep 4 pt
\label{table-RV}
\begin{tabular}{cccrc} 
\hline
\hline \noalign{\smallskip}
  \# & $JD$     & $\phi$     &  $RV~{\rm (ph)}$  &  $RV~{\rm(\ion{Fe}{ii})}$\\  
& $2450000+$ &   &  [km~s$^{-1}$]  &   [km~s$^{-1}$] \\  
\noalign{\smallskip} \hline \noalign{\smallskip}
$1$ & $1354.251$    &   $0.45$    & $-21.6 \pm 1$ & $-38.5\pm 3$ \\  
$2$ & $1654.750$     & $0.24$   & $25.2\pm 1$ &  $-65.5\pm 2$ \\ 
$3$ & $1656.761$     & $0.26$   & $24.3\pm 1$  &  $-59.3\pm 2$ \\ 
$4$ & $5337.779$    & $0.43$    &$-15.4\pm 1$ &   $\cdots$ \\ 
$5$ & $5339.840$    & $0.45$    &$-22.5\pm 2$ &   $\cdots$ \\ 
$6$ & $6455.718$    & $0.81$    &$-71.4\pm 2$ &  $\cdots$ \\ 
7 & 6698.839    & 0.07    & -4.1 &  $\cdots$ \\ 
$8$ & $6738.878$    & $0.44$    &$-19.2\pm 1$ &  $\cdots$ \\ 
$9$ & $6759.870$   &  $0.64$    & $-71.5\pm 3$ &  $\cdots$\\ 
$10$ & $6762.723$   &  $0.66$     &$-72.7\pm3$ &  $\cdots$\\ 
$11$  & $7147.775$   &  $0.25$    &  $24.2\pm 1$ & $-96.3\pm 3$ \\ 
$12$ & $7307.521$   &  $0.72$    &$-76.2\pm 2$ &  $~~26.4\pm 3$ \\ 
$13$ & $7311.494$   &  $0.76$   &$-73.6\pm1$  &  $~~24.4\pm2$  \\
$14$ & $7491.660$   & $0.43$    & $-16.4\pm1$ & $-39.8\pm2$ \\ 
$15$ & $7597.598$   &   $0.41$  & $-11.3\pm2$ &  $-43.3\pm3$ \\ 
$16$ & $7991.598$    & $0.08$    &   $-4.0\pm3$ & $-80.7\pm4$ \\ 
$17$ & $7994.600$    &  $0.10$    &  $1.2\pm1$ &  $-88.7\pm2$ \\ 
$18$ & $7998.525$    & $0.14$        &   $10.3\pm2$  &  $-85.5\pm2$ \\ 
\noalign{\smallskip} \hline \noalign{\smallskip}
\end{tabular}
\end{center}
\vspace{-0.4cm}
\end{table}

\begin{figure}[h]
\centering
\includegraphics[width=0.85\hsize,angle=-90]{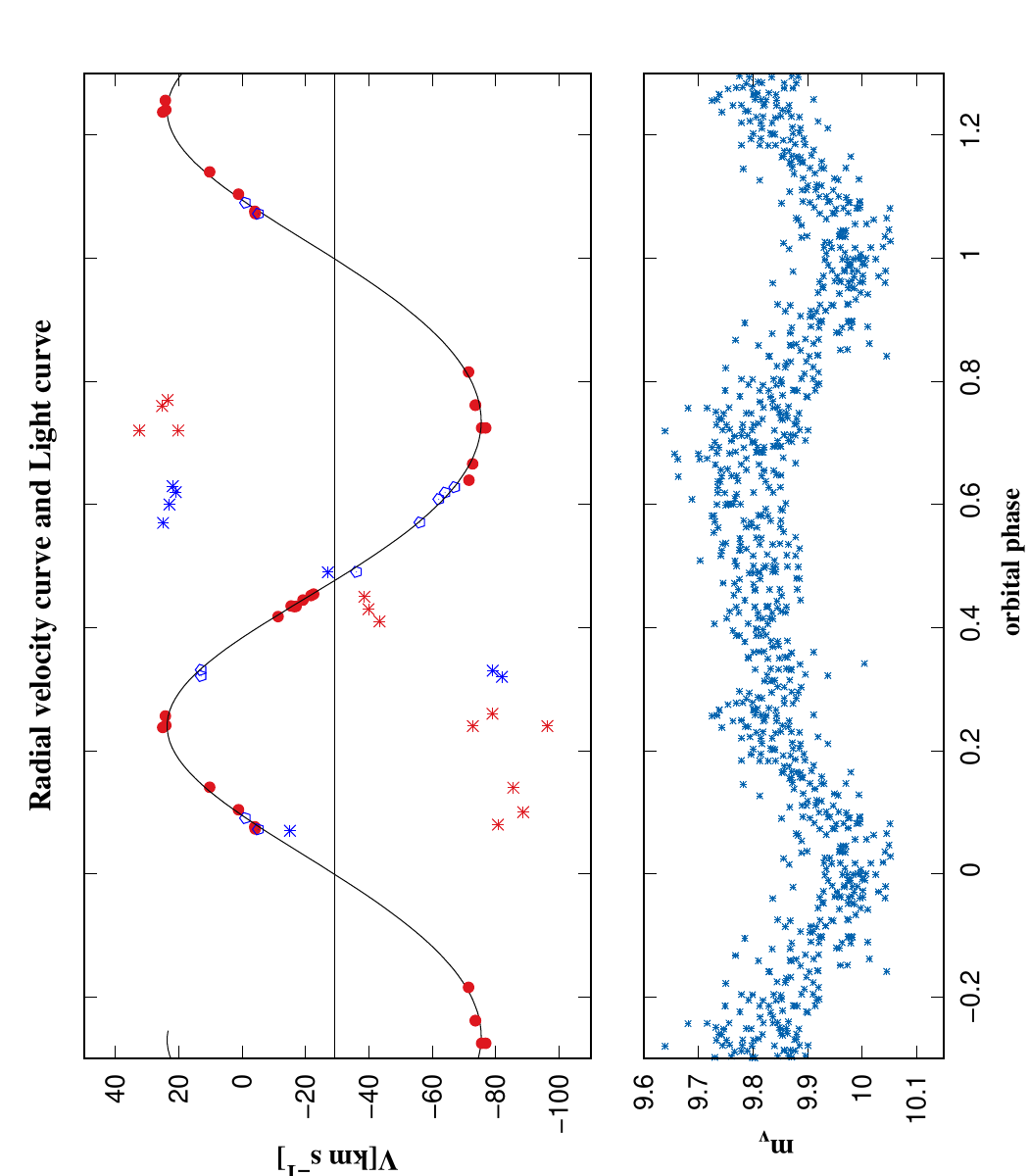}
\qquad
\caption{Radial velocity curve and ASAS-3 light curve. {\it Top panel:} Best-fitting RV curve (phase diagram) obtained with a period of 107.699 days using our data (red symbols) and those published by \citet[][blue symbols]{Miroshnichenko2003}.  Circle symbols follow the orbital motion of the F-type star (photospheric lines), and asterisks are for Fe II emission lines. The horizontal line indicates the barycentric RV of the system,  
$\gamma_0= -$29\,km\,s$^{-1}$.
{\it Bottom panel:} ASAS-3 light curve as a function of the orbital phase (see details in section \ref{ASAS}). The zero phase was calculated using the epoch $T_{\rm c}$, when the cool component is located between the observer and
the hot component.}
\label{fig-orbits}
\end{figure}
\begin{table*}[h]
\begin{center}
\caption{Orbital parameters of the binary system. The term $T_{\rm c}$ corresponds to a minimum brightness epoch observed in the ASAS-3 light curve and defines a zero phase.
\label{table-orbit}}
\begin{tabular}{cccccc} 
\hline
\hline
\noalign{\smallskip}
  $P$       &  $K$         & $\gamma_0$   &     $e$     & $T_{0} $  & $T_{\rm c}$ \cr
  [days]  & [km~s$^{-1}$] & [km~s$^{-1}$] &                &  [JD-2450000]  &    [JD-2450000] \cr 
  \noalign{\smallskip}
\hline
\noalign{\smallskip}        
$107.699\pm0.005$ &  $49.55\pm0.42$  & $-29.23\pm0.30$ & $0.070\pm0.007$  & $1333.25\pm6.46$   &  $2813.823$ \cr         
\noalign{\smallskip}
\hline
\end{tabular}
\end{center}
\end{table*}

\begin{figure}[h!]
\begin{center}
\includegraphics[width=0.33\hsize,angle=-90]{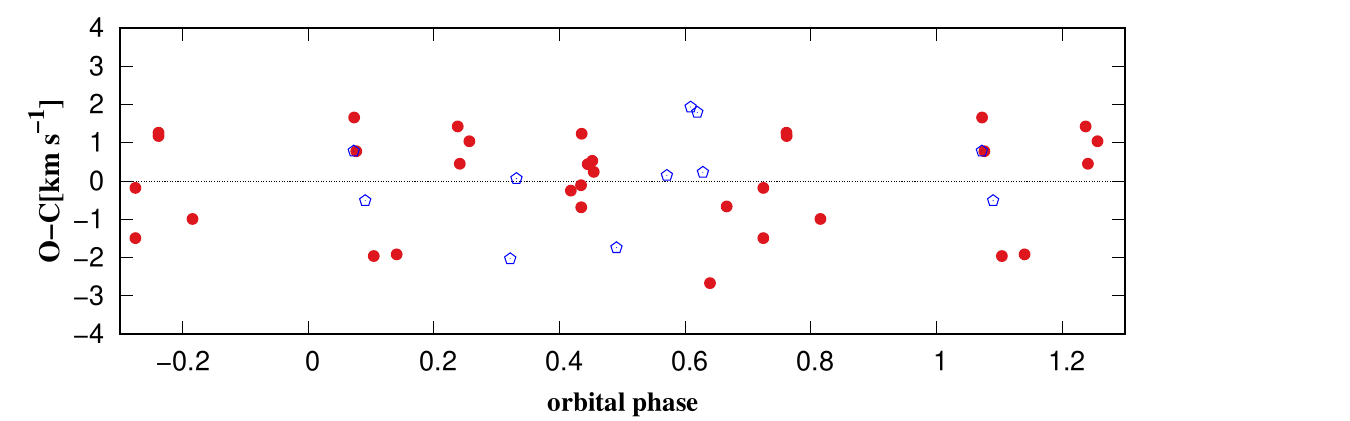}
\qquad
\caption{Residuals of the RV curve. The symbol description is the same as Fig.~\ref{fig-orbits}.}
\label{O-C}
\end{center}
\end{figure}

We used the RVs from the metallic lines to improve the binary system's orbital solution. These values were combined with data published by \citet{Miroshnichenko2003} to calculate the resulting RV curve, shown in Fig.~\ref{fig-orbits}. The best-fitting model was obtained by minimising $\sum_{i=1}^{N} w_{\rm i} (O-C)^2$, where O and C are the observed and calculated RV values, respectively. We used the same weight, $w_{\rm i}$, for all the data. The residual values are plotted in Figure~\ref{O-C}. It can be seen that the absolute errors in the RVs do not exceed $2.7$~km~s$^{-1}$.

The improved parameters of the orbital solution are listed in Table~\ref{table-orbit}. We derived a period of $P=107.699\pm0.005$ days \citep[very similar to that reported in][]{Maravelias2018, Nodyarov2024} and an RV semi-amplitude $K = 49.55\pm0.42$~km~s$^{-1}$. The rest of the parameters listed in the table are the barycentric RV of the system ($\gamma_0$), the eccentricity ($e$), and the time of periastron passage ($T_0$), with their respective errors. $T_{\rm c}$ corresponds to a minimum brightness epoch observed in the ASAS-3 light curve and defines a zero phase.

\vspace{-0.05cm}
The average of the RV measurements of the emission lines \ion{Fe}{ii} $\lambda\lambda 5991, 6084$\,\AA, also shown in Fig.~\ref{fig-orbits} (top panel), exhibits anti-phased behaviour when compared to the photospheric lines of the cool component, supporting the results of \citet{Miroshnichenko2003} and \citet{Nodyarov2022}. As all the absorption lines used for the analysis correspond to the late-type star, the presence and RV behaviour of emission lines suggest that they arise in a disc around the hot component. 
\vspace{-0.05cm}

The mass function of our system, $f(m)$, can be obtained from
\vspace{-0.1cm}
\begin{equation}
  f(m) = {\frac{1}{2 \pi G} }P K_{1}^{3} \left(1-e^{2}\right)^{1.5} = \frac{(\mathfrak{M}_2 \sin i)^3}{(\mathfrak{M}_1 + \mathfrak{M}_2)^2},   
\label{Eq:m}
\end{equation}

\noindent where $K_1$ and $\mathfrak{M}_1$ are the RV semi-amplitude and mass of the cool companion (the F-type star, since it is the main contributor to the optical spectrum) respectively, $\mathfrak{M}_2$ is the mass of the emission-line B-type star, $i$ the orbit inclination, $G$ the gravitational constant, $P$ the orbital period, and $e$ the eccentricity. With the values adopted in Table~\ref{table-orbit}, we find $f(m) = 1.357 \pm 0.002~ M_{\sun}$.

Due to the uncertainty in the inclination angle $i$ of the orbital plane, the mass $\mathfrak{M}_2$ of the emission-line object becomes indeterminate. Nevertheless, we can estimate a range of mass values considering that the mass $\mathfrak{M}_2$ is $q$ times the mass $\mathfrak{M}_1$ of the F-type star (donor star). Thus, the mass function, $f(m)$,  is now expressed in terms of the parameter $q$ as 
\begin{equation}
   f(q)= \mathfrak{M}_1~\sin^{3}i ~\frac{q^{3}}{(1+q)^{2}}. 
   \label{Eq:q}
\end{equation}

The mass function depends on the mass of the donor component and the inclination angle $i$. Moreover, as the RV curve traced by the emission lines has almost the same amplitude as that of the late-type star ($K_1 \sim K_2$), this implies $q\sim 1$. A lower amplitude would lead to a higher stellar mass. Therefore, the minimum mass expected for the emission-line star companion is when $q=1$ (see discussion in Sect.~\ref{Disc}).

\subsection{Light curve analysis}
\label{ASAS}

\begin{figure}[h]
\vspace{-0.8cm}
\begin{center}
\includegraphics[width=10.5cm,angle=0]{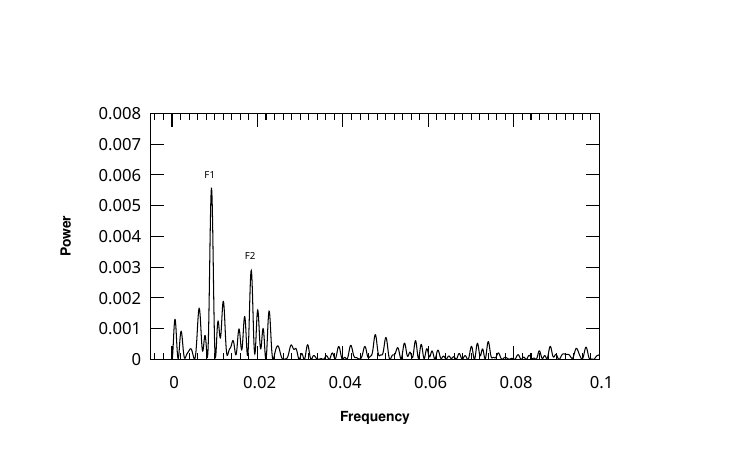}
\caption{Periodogram of the ASAS-3 light curve showing two dominant frequencies, $F1= 0.0092772$~d$^{-1}$ ($P=107.79$~d) and $F2= 0.018579$~d$^{-1}$ ($53.82$~d).}
\label{periodogram}
\end{center}
\end{figure}

The light curve of \object{HD\,327083} is available in the ASAS Photometric V-band Catalogue \citep{Pojmanski1997}. 
Based on the discrete Fourier transform algorithm and a least-squares fitting, we performed a time series analysis of the ASAS-3 light curve with the Period04 software \citep{Lenz2014}. The power spectrum is shown in Fig.~\ref{periodogram}. Two significant frequencies are detected. The first frequency yields a period of $P=107.79$~d, and the second one identifies its second harmonic, $P=53.82$~d. The first period is very close to the one derived from our orbital solution ($ P=107.699$~d). 
Figure~\ref{fig-orbits} (bottom panel) illustrates the light curve folded with the orbital phase, adopting the period given in Table~\ref{table-orbit}. A deep and broad minimum occurs when the F-type star is located between the observer and the B-type star ($\phi=0$). A shallow minimum is observed at $\phi=0.5$. 
According to the model by \citet[][in their Fig.~4]{Nodyarov2024}, the characteristic shape of a nearly Roche-lobe-filling star combined with reflection effects caused by illumination from the hot star and its accretion disc leads to a double-wave modulation in the light curve over the orbital period, as seen in Fig.~\ref{fig-orbits} (bottom panel).

\begin{figure}[h]
\vspace{-0.4cm}
\includegraphics[width=6.5cm,angle=-90]{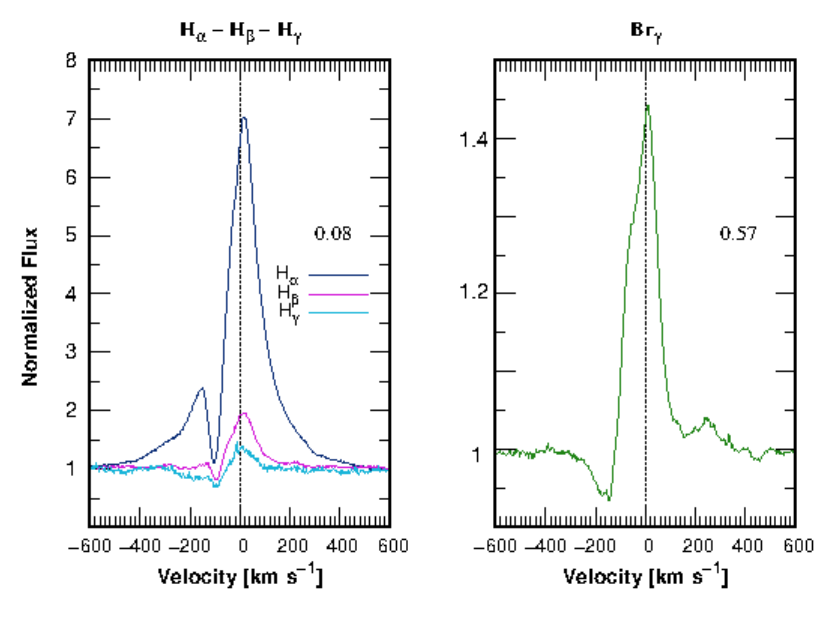}
\caption{Optical and near-IR H lines. {\it Left panel:} Line profiles of H$\alpha$, H$\beta$, and H$\gamma$ at $\phi = 0.08$. {\it Right panel:} P~Cygni profile of the Br$\gamma$ line at $\phi=0.57$.
\label{H}}
\end{figure}

\begin{figure*}[h!]
\begin{center}
\includegraphics[width=5.8cm,angle=-90]{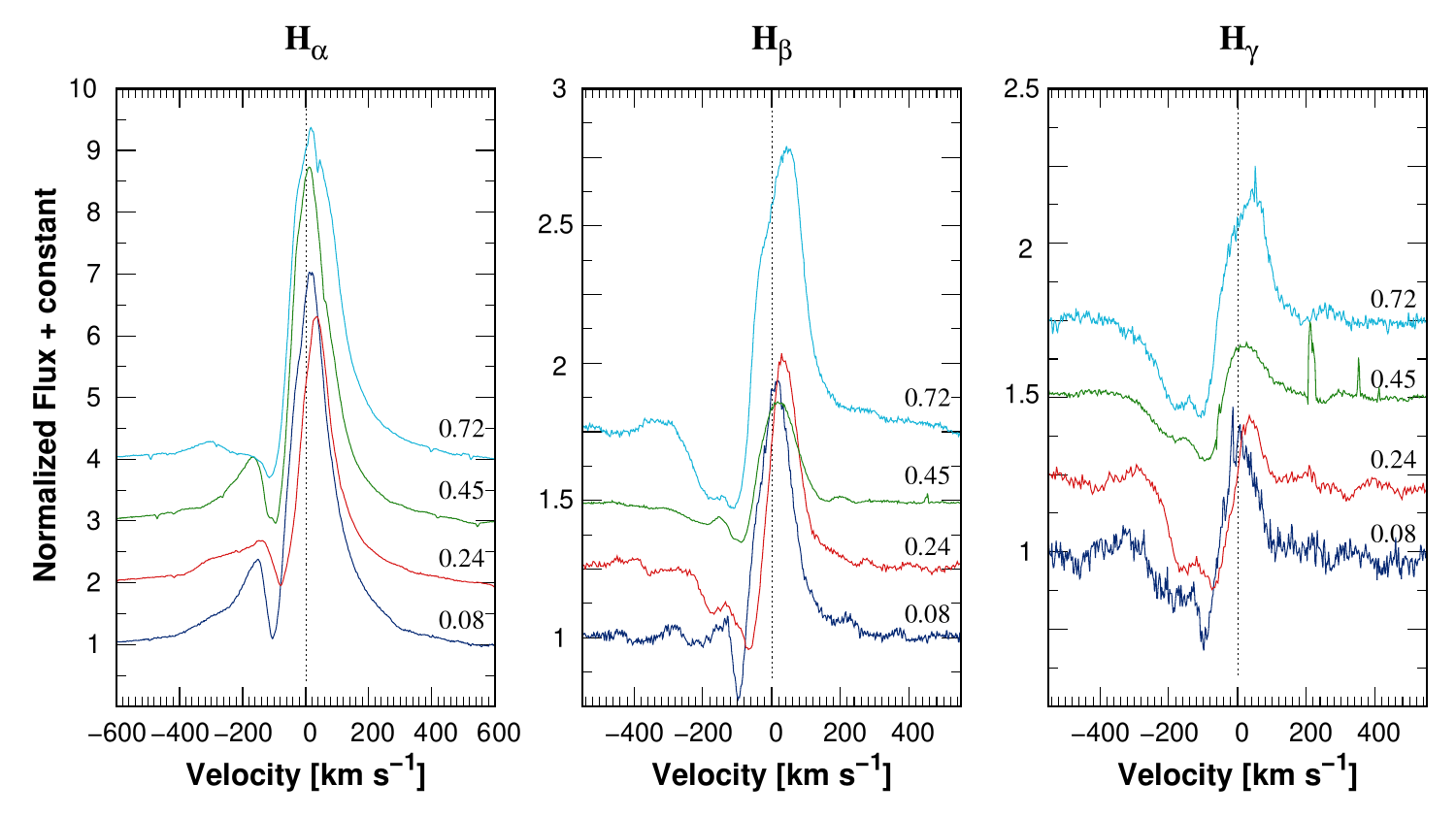}
\hspace{-0.2cm}
\includegraphics[width=5.8cm,angle=-90]{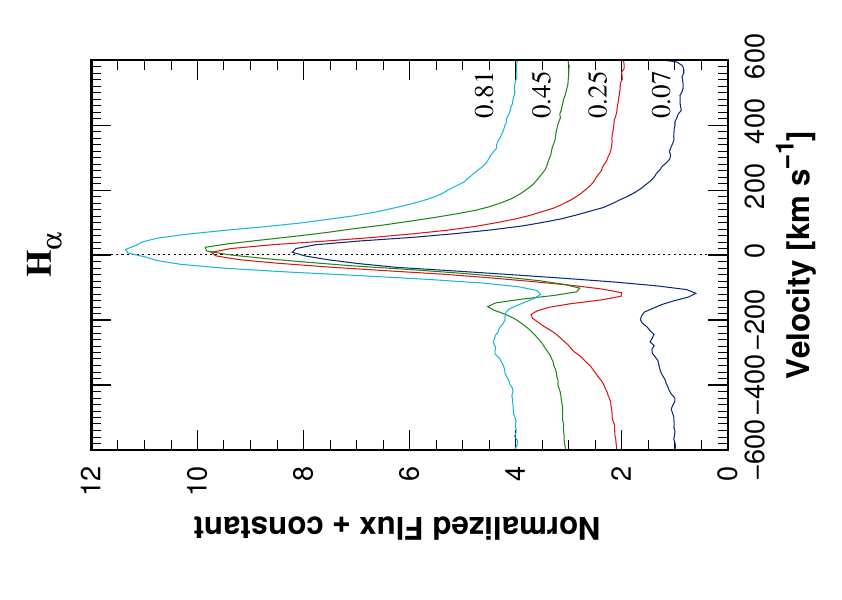}
\caption{Spectra from FEROS showing changes in the H$\alpha$ (first panel), H$\beta$ (second panel), and H$\gamma$ (third panel) lines with the orbital phase. Variations of the H$\alpha$ line were observed with REOSC (fourth panel). Spectra have been shifted vertically to facilitate comparison.  \label{H-lines}}
\end{center}
\end{figure*}

\begin{figure}[h!]
\hspace{0.5cm}\includegraphics[width=5.9cm,angle=-90]{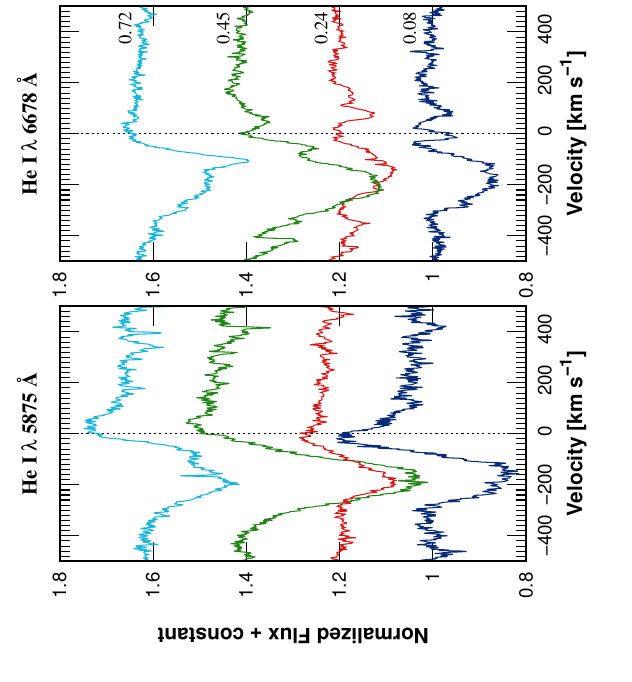}\vspace{-0.5cm}
\begin{center}
\includegraphics[width=5.9cm,angle=-90]{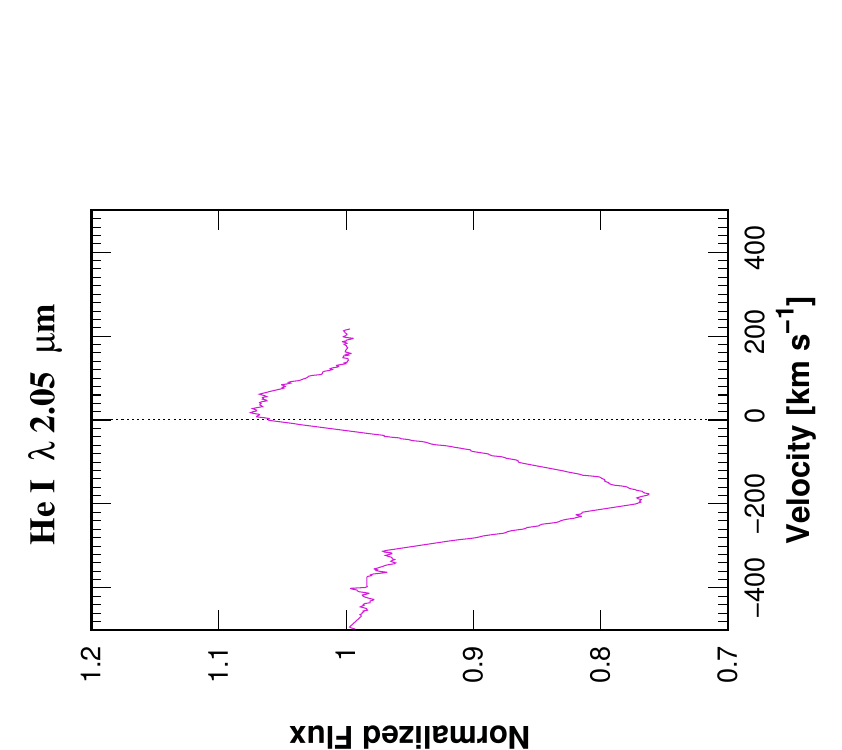}
\end{center}
\caption{Helium line profiles. {\it Top panel:} Variations of the \ion{He}{i} lines (taken with FEROS) with the orbital phase. Spectra have been shifted vertically to facilitate comparison. {\it Bottom panel}: Infrared line of \ion{He}{i}~$\lambda2.05~\mu$m observed with IGRINS at the orbital phase $\phi=0.57$.}
\label{HeI}
\end{figure}

\subsection{Optical and near-IR spectra}
\label{sect:opt-spectra}

In this section, we briefly describe the most prominent spectral lines. We then examine their time evolution over the orbital period.

a) H lines: The H$\alpha$ and H$\beta$ lines display a blue-shifted absorption with emission components at both sides \citep[a P~Cygni type V profile,][]{Beals1953} during almost the whole orbital cycle (see Fig.~\ref{H}, left panel). At $\phi=0.08$, the H$\alpha$ emission is $7$  times the continuum level, while the H$\beta$ emission line is only twice. The RV of the absorption component of H$\alpha$ and H$\beta$ cores are respectively $-70$~km\,s$^{-1}$ and $-66$~km\,s$^{-1}$ (corrected by the barycentric RV of the orbital motion). The H$\gamma$ line presents, in addition to a narrow absorption component at $-43$~km\,s$^{-1}$, a typical P~Cygni profile with an extended blue-shifted absorption with a RV of $-276$~km\,s$^{-1}$ at the blue edge. The Br$\gamma$ line (shown on the right panel of Fig.~\ref{H}) displays a P~Cygni profile with a barycentric corrected RV of $-139.7$ km~s$^{-1}$ at the blueward-absorption core when the orbital phase is $0.57$. The emission component is somewhat asymmetric. The rest of the lines of the Brackett series are formed by the overlapping of \ion{H}{} absorption lines of the F-type star and emission features from the hot circumstellar envelope, giving a `fake' appearance of a P~Cygni profile.

\begin{figure*}[h!]
\hspace{-0.4cm}
\includegraphics[width=5.9cm,angle=-90]{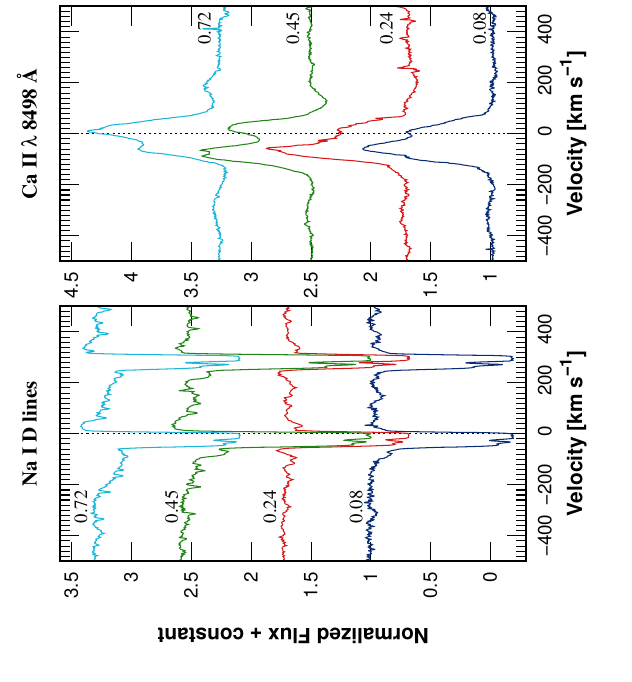}\hspace{-0.3cm}
\includegraphics[width=5.9cm,angle=-90]{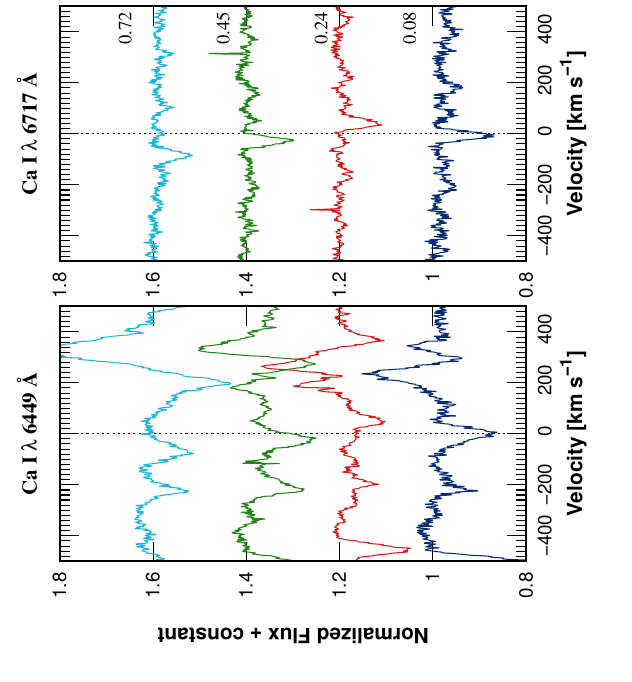}
\hspace{-0.3cm}
\includegraphics[width=5.9cm,angle=-90]{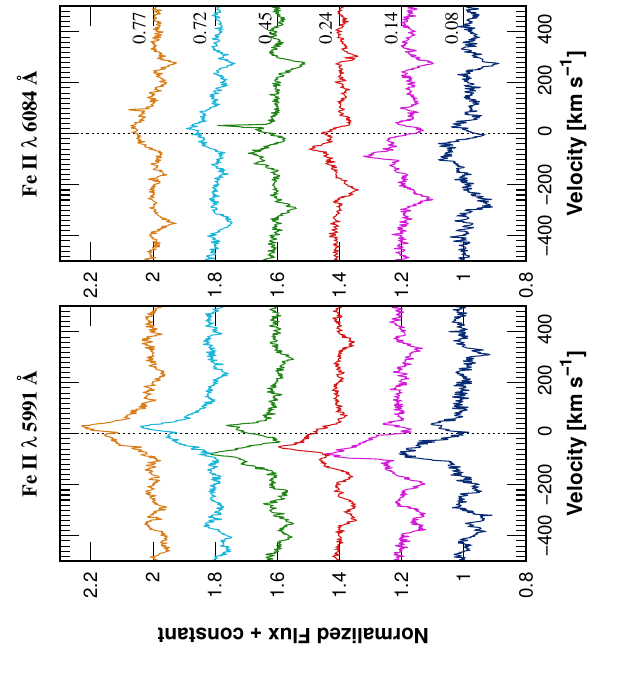}
\caption{Line variations with the orbital phase. {\it Left panels:} \ion{Na}{i} D1 and D2 lines and \ion{Ca}{ii} $\lambda 8498$~\AA\, line. {\it Middle panels:} \ion{Ca}{i} lines.
{\it Right panels:} \ion{Fe}{ii}~(46). These lines trace the motion of the hot component.
Spectra have been shifted vertically to facilitate comparison.}
\label{NaI}
\end{figure*}

Figure~\ref{H-lines} shows the behaviour of the H$\alpha$, H$\beta$, and H$\gamma$ line profiles seen in FEROS spectra with the orbital phase. The REOSC spectra, taken almost at the same orbital phase, exhibit the same response (right panel).  
The Balmer lines change their intensity and widths with the orbital phase, being wider between phases $0.72$ and $0.81$. Close to the quadrature phases ($0.24$ and $0.75$), the blue emission component fades and develops an extended wing. The relative intensity of the absorption and emission components also changes along the orbital motion; the lowest intensity is observed between $\phi=0.24$ and $\phi=0.45$. A weakness in the intensity of the P~Cygni emission is also found in the B\&C spectrum taken at $\phi=0.37$. 

b) \ion{He}{i} lines: The line of \ion{He}{i} $\lambda$4471 \AA\, is highly contaminated by metallic lines from the F-type star (see Fig.~\ref{optical_feros}). In contrast, the \ion{He}{i} $\lambda 5875$~\AA\, and \ion{He}{i} $\lambda 6678$~\AA\, lines display clear P~Cygni profiles (see Fig.~\ref{HeI}, top panel). The P~Cygni profiles also change significantly with the orbital phase. At $\phi=0.45$, the absorption component is deeper and wider than in other phases. The absorption core is blue-shifted with barycentric velocities ranging from $-182$~km\,s$^{-1}$ to $-235$~km\,s$^{-1}$. At $\phi=0.72$, we estimate an RV of $-366$~km\,s$^{-1}$ at the blue edge of the P~Cygni absorption line, corrected by the barycentric RV, which might be interpreted as due to an outflowing polar wind. The IR \ion{He}{i} line at $2.05~\mu$m (bottom panel of Fig.~\ref{HeI}) also shows a broad P~Cygni profile with velocities of about $-162$~km\,s$^{-1}$ and $-378$~km\,s$^{-1}$ for the core of the P~Cygni absorption component and the blue-shifted edge, respectively.    

c) \ion{Na}{i}, \ion{Ca}{i}, \ion{Ca}{ii}, and \ion{Mg}{ii} lines: The D$_1$ and D$_2$  \ion{Na}{i} lines also show changes with the orbital phase (see Fig.~\ref{NaI}, left panel). P~Cygni-type profiles are present between phases $0.45$ and $0.72$, with blue extended wings. Line absorption components with velocities of $-20$~km\,s$^{-1}$, $-24$~km\,s$^{-1}$, $-135$~km\,s$^{-1}$, and $-149$~km\,s$^{-1}$ are present. An inverse P~Cygni feature is noted at $\phi=0.24$. We also notice that the profiles of  \ion{Na}{i}~D-lines exhibit short-term variations, both in intensity and shape. 

\begin{figure}[h!]
\centering
\includegraphics[width=5.9cm,angle=-90]{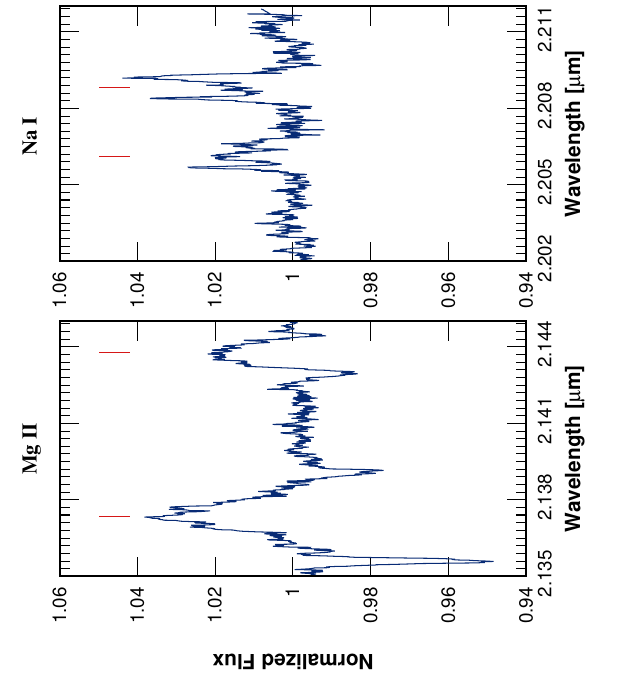}
\caption{Emission lines of \ion{Mg}{ii} and \ion{Na}{i} observed in the K-band spectrum. The absorption features belong to the cool star.}
\label{Mg-Na}
\end{figure}

The \ion{Ca}{ii} triplet is also seen in emission and displays variation with the orbital phase. The Fig.~\ref{NaI} (second panel) shows intensity and shape variations of the line \ion{Ca}{ii}~$\lambda 8498$~\AA. At phase $\phi=0.45$, the line shows a double-peaked emission with a separation of $-57$~km\,s$^{-1}$. Instead, the \ion{Ca}{i} lines are in absorption (middle panels). They show RV and intensity changes with the orbital motion of the F-type star.
The K-band spectrum, shown in Fig.~\ref{Mg-Na}, displays broad and weak emission lines of \ion{Na}{i} $\lambda\lambda 2.206, 2.209~\mu$m  (right panel) and \ion{Mg}{ii} $\lambda\lambda2.137, 2.144~\mu$m (left panel), the former showing a double-peaked profile. The absorption features seen in the figure belong to the cool star.

d) \ion{Fe}{ii} lines: The \ion{Fe}{ii} lines ($\lambda\lambda 5991, 6084$\,\AA) from multiplet~$46$ are observed in emission (see right panels of Fig.~\ref{NaI}). The RV variations of some lines follow the motion of the gaseous envelope around the hot component. Other lines show RVs close to the barycentric RV of the system. The former is probably formed in the inner region of the accretion disc around the primary star. The low RVs observed in the other \ion{Fe}{ii} lines suggest a circumbinary origin. Forbidden lines of \ion{Fe}{ii} are absent.

e) The CO band emission: The CO first-overtone bands arise in the K band at wavelengths longer than $2.293~\mu$m. The entire band structure has been observed with IGRINS, whereas the observations with Phoenix and CRIRES covered only portions around the first and second band heads (see Table~\ref{table:2}). The second-overtone bands that arise in the H band in the wavelength region 1.555-1.680 $\mu$m are covered by the IGRINS spectrum. 

\begin{figure*}[h!]
\begin{center}
\includegraphics[width=0.96\hsize,angle=0]{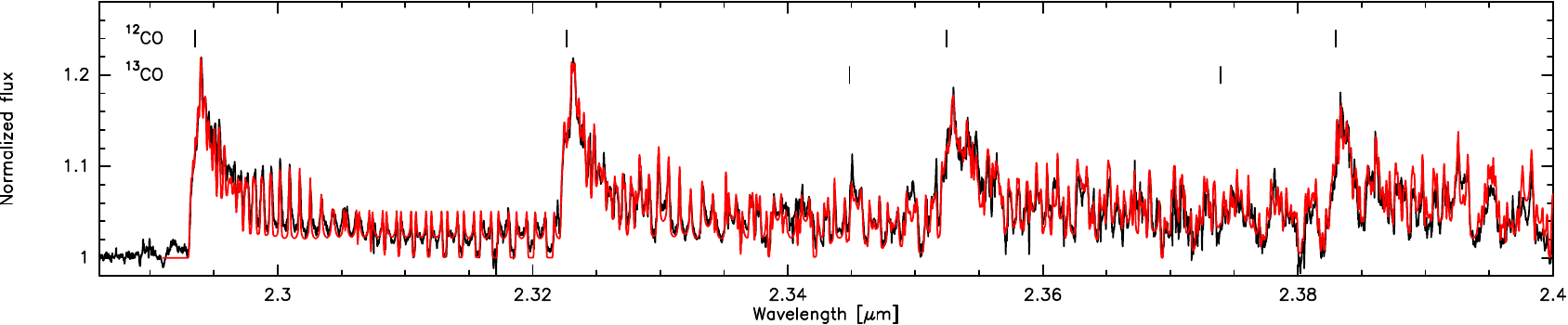}
 \caption{Portion of the normalised IGRINS K-band spectrum covering the CO first-overtone bands (solid black line). The best-fitting CO model is shown in red. The tics indicate the positions of $^{12}$CO and $^{13}$CO band heads.}
 \label{fig-igrins_CO}
 \end{center}
\end{figure*}

To analyse the CO band emission, we focus first on the IGRINS K-band spectrum, shown in Fig.~\ref{fig-igrins_CO}. Its wide wavelength coverage, up to $2.45~\mu$m, encompasses five band heads of $^{12}$CO and four band heads of $^{13}$CO, which is essential to constrain the physical parameters of the molecular gas precisely. The high resolution of the spectrum allows for identification of individual ro-vibrational transitions, especially bluewards of the second band head. These lines display double-peaked profiles, which indicate rotational broadening, in agreement with the findings of  \citet{Wheelwright2012b} that the disc harbouring the CO gas revolves around the binary system on Keplerian orbits. 
Furthermore, \citet{Wheelwright2012a} detected the presence of $^{13}$CO based on their low-resolution AMBER data.

We used the code of \citet{Kraus2000}, developed to compute $^{12}$CO band emission from a Keplerian rotating disc under LTE conditions. This code has been extended by \citet{Kraus2009} and \citet{Oksala2013} to include emissions from the isotopic molecule $^{13}$CO. We implemented updated values for the energy levels and Einstein transition coefficients from \citet{2015ApJS..216...15L}.
This update was also necessary to compute the CO emission of the second overtone bands arising in the H band precisely. 

\citet{Kraus2009} also showed that the CO molecular band emission intensity increases with both the temperature and column density. Therefore, the observed emission typically traces the hottest and densest region, which can be identified as the inner rim of the CO molecular disc. Under such conditions, the assumption of LTE is well justified, and it also allows us to restrict our calculations to a single ring of gas with constant temperature and column density.

\begin{figure}[h!]
\centering
\includegraphics[width=0.7\hsize,angle=-90]{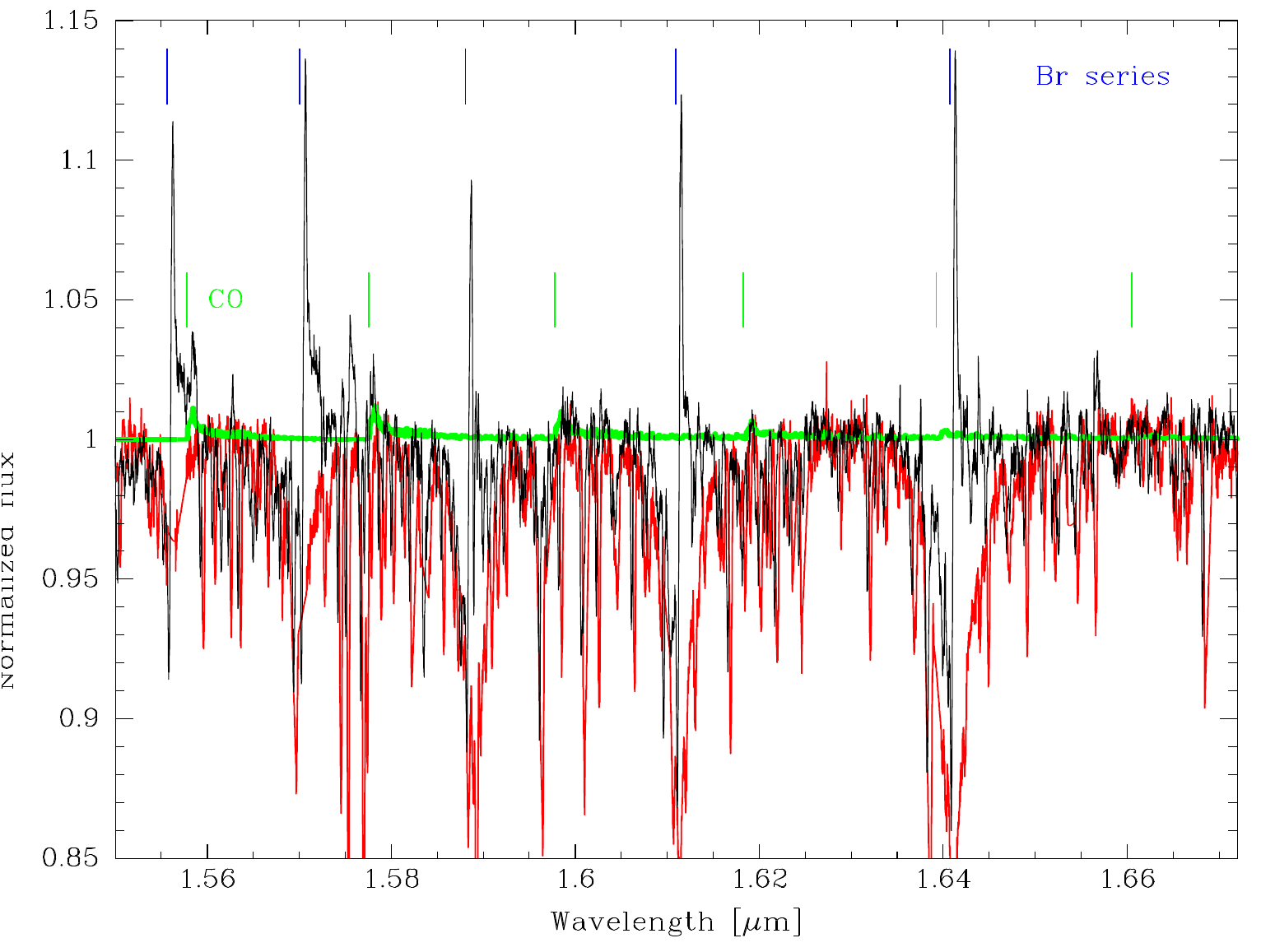}
 \caption{Second-overtone bandheads of the CO molecule. The complex features seen in the spectrum (solid black line) combine the CO molecular emission pattern and the Brackett lines in absorption arising from the photosphere of the cool companion. A standard F-type star's spectrum (solid red line) and the CO model (solid green line) are plotted to facilitate comparison.}
 \label{fig:2daCO}
\end{figure}

The CO parameters of the best-fitting model to the IGRINS spectrum are
given in Table\,\ref{tab-CO}. 
For the computation of the rotational velocity, we adopted a ring inclination angle of $48.5\degr$ \citep{Wheelwright2012a}. The model spectra are shown to facilitate comparison with the IGRINS observations in the K band (Fig.~\ref{fig-igrins_CO}) and H band (Fig.~ \ref{fig:2daCO}). The latter is very faint and significantly blended with the intense photospheric \ion{H}{i} absorption lines of the cool companion.
From our model, we also found considerable enrichment of the molecular gas with $^{13}$CO, in agreement with its origin from the surface of an evolved star.

\begin{table}[h]
\caption{Model parameters for the CO ring.}
\label{tab-CO}
 \tabcolsep 1.0pt
\begin{tabular}{ccccccc}
         \hline
         \hline
          \noalign{\smallskip}
  $N_{\rm CO}$ & $T_{\rm CO}$ && $i$     & $v_{\rm rot}$  & $v_{\rm los}$ & \small{$^{12}$CO/$^{13}$CO} \\
  $ $ [cm$^{-2}$] &      [K]     && [\degr] & [km\,s$^{-1}$] & [km\,s$^{-1}$] &  \\
         \noalign{\smallskip}
         \hline
         \noalign{\smallskip}
 $(3\pm 0.5) \times 10^{21}$  & $2300\pm 100$ && $48.5$ & $75\pm 1$ &$55.5\pm 1.0$ & $20\pm 5$  \\
 \noalign{\smallskip}
         \hline
  \end{tabular}
  \tablefoot{The columns list the CO column density and temperature of the emitting region, as well as the inclination of the CO disc \citep[taken from][]{Wheelwright2012a}, the  Keplerian velocity, the rotational velocity projected to the line of sight, and the $^{12}$CO/$^{13}$CO isotopic ratio.}
\end{table}

\begin{figure}[h]
\centering
\includegraphics[width=0.95\hsize,angle=0]{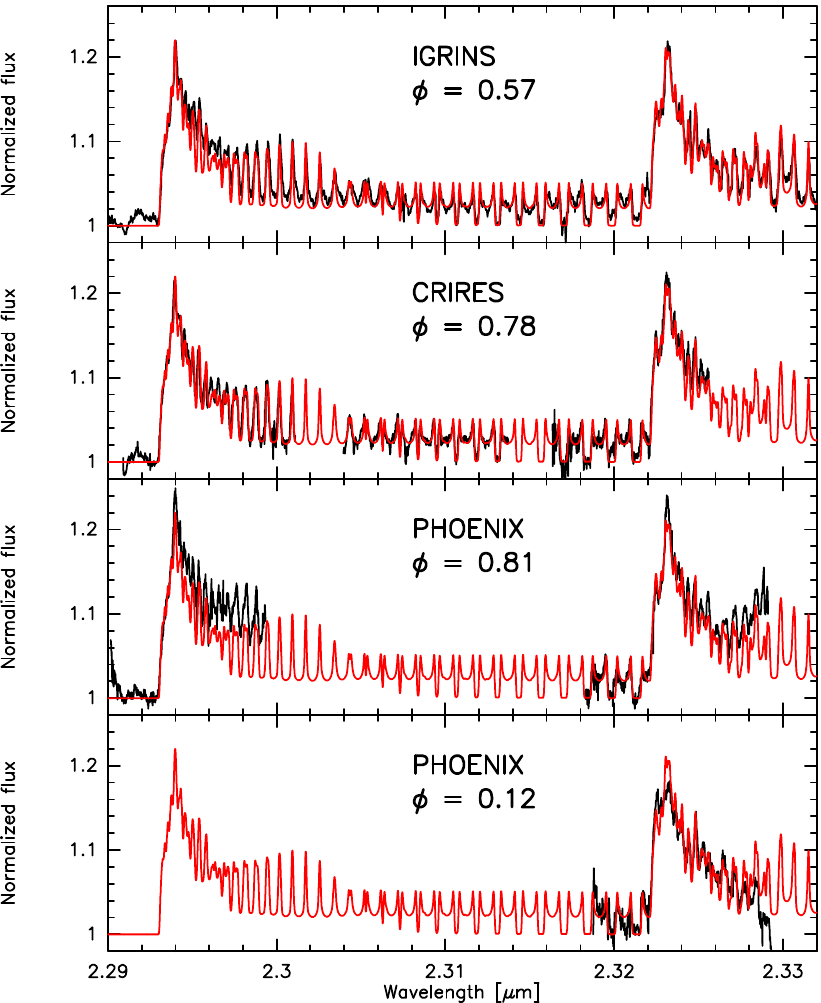}
\caption{Spectra of CO taken with various instruments at different orbital phases (black 
lines). For reference, the best-fitting model (red lines) to the IGRINS spectrum (top panel) is overplotted in all panels. A local density enhancement at $\phi=0.81$ can be observed as well as a change in the blue shoulder of the second CO band head at $\phi=0.12$.}
\label{CO_phase}
\end{figure}

\begin{figure}[h]
\centering
\includegraphics[width=0.6\hsize,angle=-90]{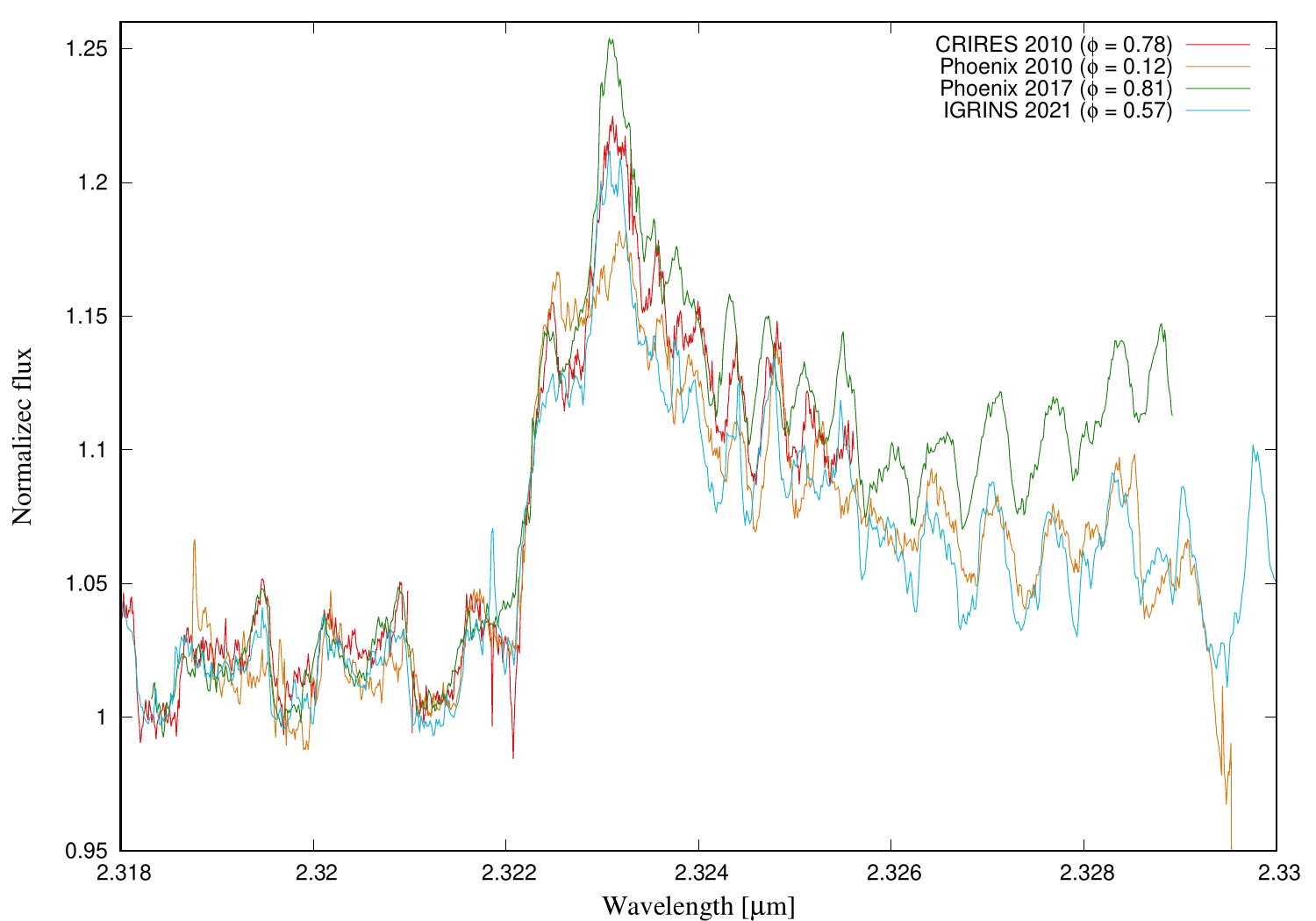}
 \caption{Variation in the intensity and shape of the second CO band head with the orbital phase.}
 \label{fig:2010-2021}
\end{figure}

In addition, we reanalysed the CO spectra taken with CRIRES in 2010. Both \citet{Wheelwright2012b} and \citet{Maravelias2018} studied only the first-overtone CO band head 
arising in the spectral range $2.276-2.326~\mu$m. 
From this, one can only derive the kinematics of the CO gas with high accuracy, but not the other parameters (temperature and column density).
However, the CRIRES spectra cover four individual short wavelength portions,  three of them located in regions of the CO band emission (as shown in the second panel of Fig.~\ref{CO_phase} with a black solid line). We found that the same parameters that fit the IGRINS spectrum also reproduce the CRIRES spectral pieces very well (shown in red solid line in Fig.~\ref{CO_phase}) despite the time difference of more than eleven years between the two datasets and their different orbital phases.

The situation changes when inspecting the two sets of Phoenix spectra, one taken at orbital phase 0.81 in 2017 (third panel in Fig.~\ref{CO_phase}) and the other one taken at phase $\phi = 0.12$ in 2010 (fourth panel). These two spectra cannot fit equally well with the CO model (shown in red lines that were added just for comparison purposes). The most striking difference is that the entire observed emission is more intense in phase 0.81. In addition, the individually resolved lines display
asymmetric line profiles with the red peak more intense than the blue one.
At phase 0.12, it seems to be the opposite. The CO band emission is less intense, and the individual line profiles have a slightly higher blue peak. Unfortunately, at this phase, the first band head was not observed. Still, the difference with the other observational sets is obvious, as can also be seen more clearly in Fig.~\ref{fig:2010-2021}, where we compare the shape and intensity of the second band head observed at different orbital phases. Noteworthy, the strongest deviations in the spectrum are seen at phases 0.81 and 0.12, which mark more or less the beginning and the end of the broad minimum in the light curve, which might suggest that the late-type star hides part of the CO ring.

Finally, it is important to stress that the RV values derived from the CO band heads are between -8~km~s$^{-1}$ and 3~km~s$^{-1}$  relative to the barycentre of the system. Notably, the IGRINS spectra were taken at the orbital phase $\phi = 0.57$, confirming that the ring of molecular gas is circumbinary.

\subsection{VLTI/MIDI}

\begin{figure*}[h]
\hspace{-0.5cm}
\includegraphics[width=6.5cm]{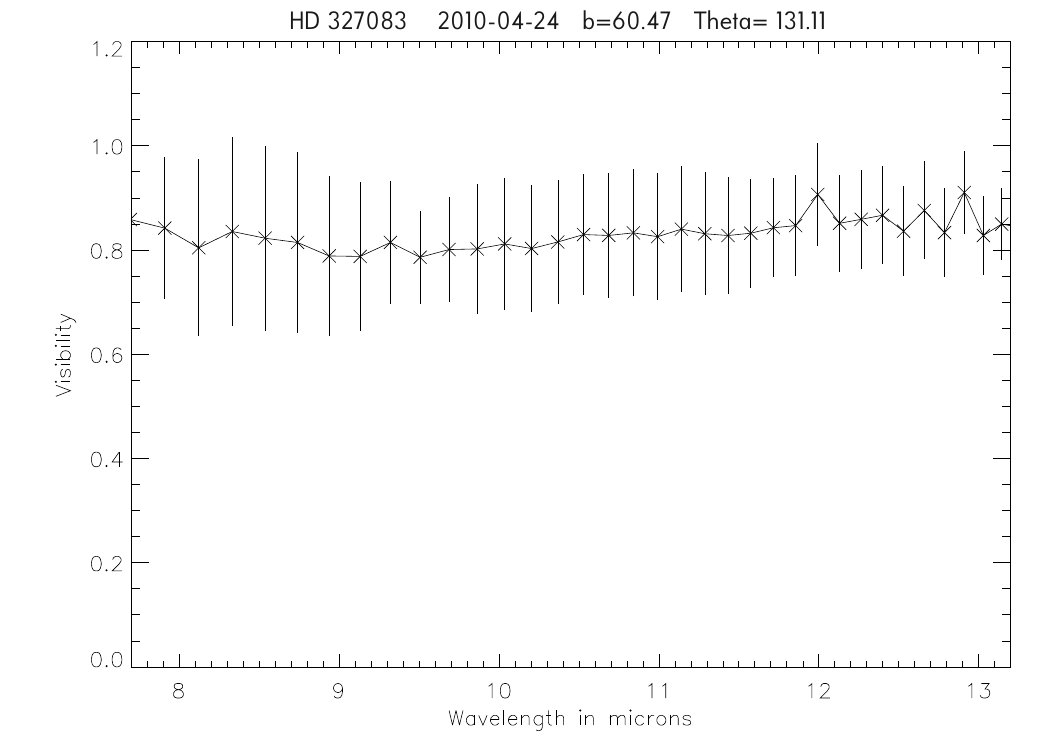}\hspace{-0.3cm}
\includegraphics[width=6.5cm]{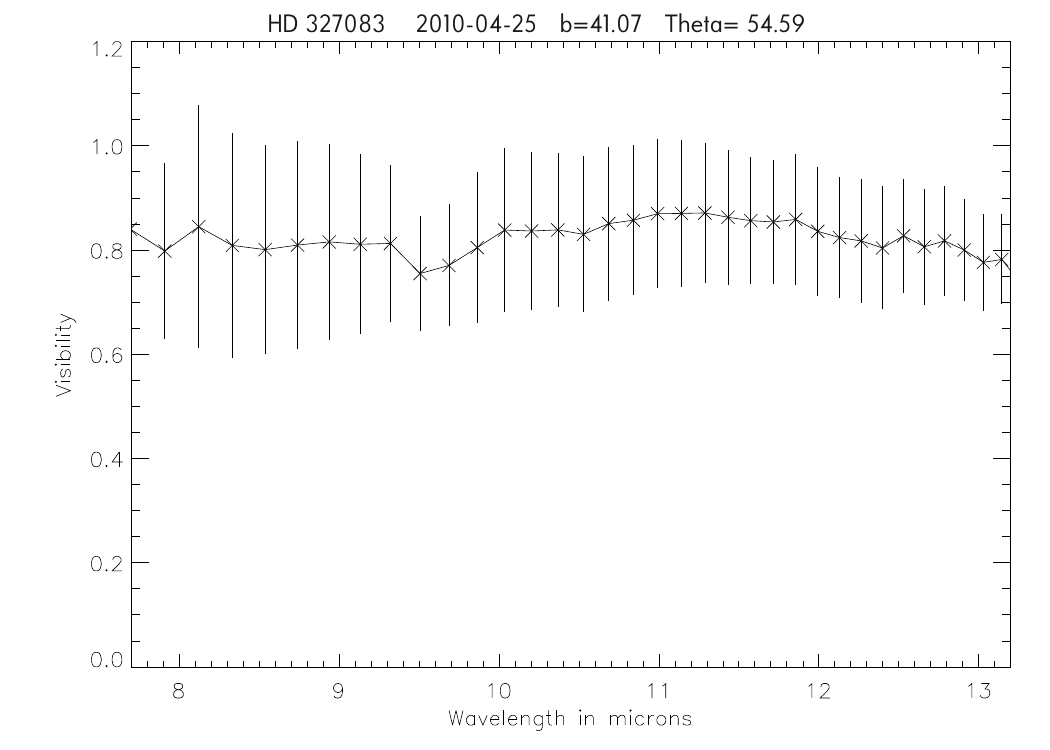} \hspace{-0.3cm}
\includegraphics[width=6.5cm]{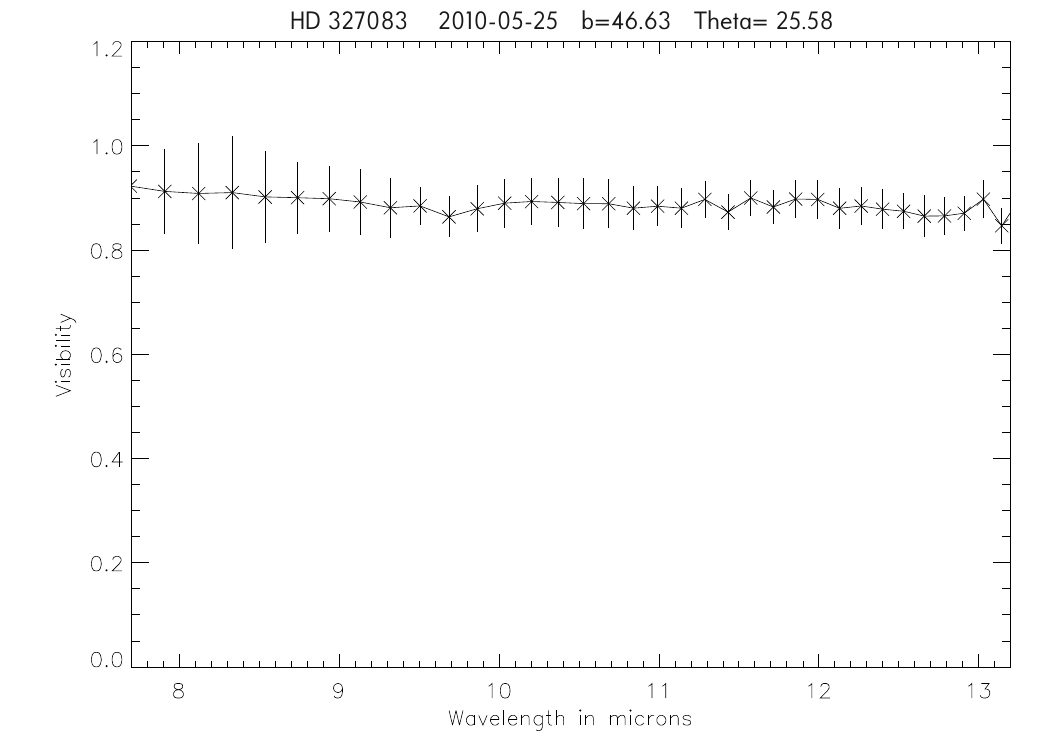}
\caption{VLTI/MIDI visibility curves for \object{HD\,327083} during April--May 2010. The curves are almost flat, with values near $0.8$.}
\label{VisiHD-32}
\end{figure*}

\begin{figure}[h]
\includegraphics[width=0.93\hsize]{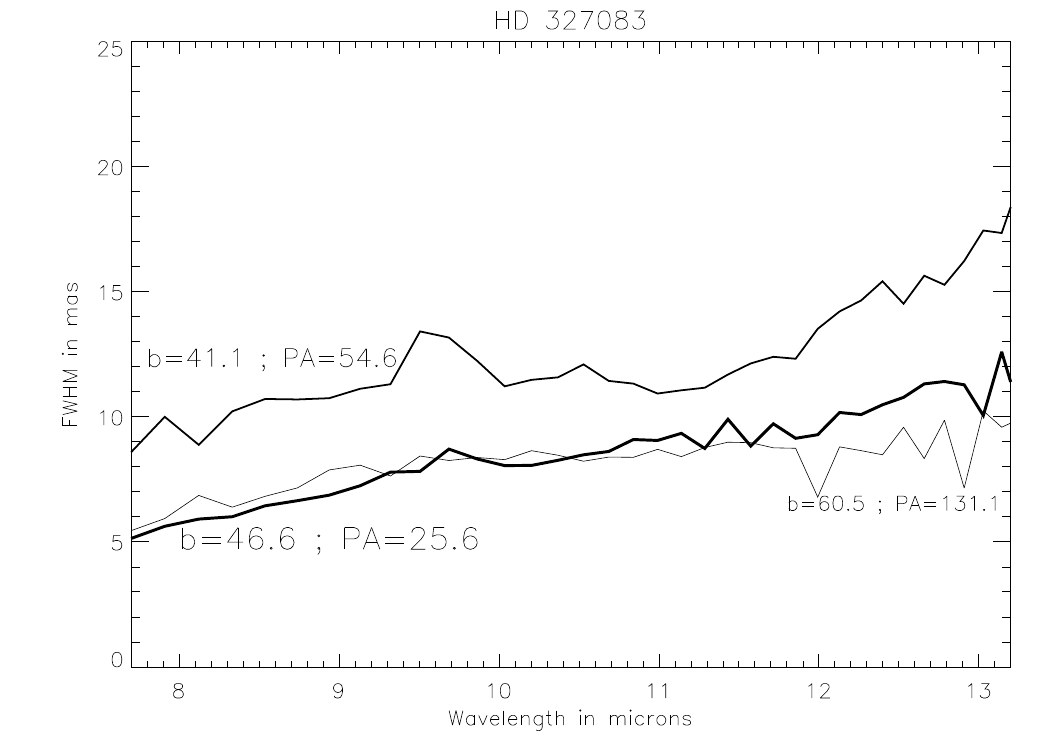}
\caption{VLTI/MIDI Gaussian fits for each visibility curve as a function of wavelength. The extent of the dust regions varies with wavelength.}\label{Gauss_HD32} 
\end{figure}

Figure~\ref{VisiHD-32} displays the MIDI visibility curves of \object{HD\,327083} for the three selected baselines and the given set of position angles in the sky. The visibility curves are almost flat, with values near~$0.8$. The error bars denote the total calibration uncertainty, including both random and systematic errors. We also calculated the full width half maximum (FWHM) from the observed visibility curves, assuming a $1-$D Gaussian model for the envelope intensity distribution \citep{Leinert2004}, as shown in Fig.~\ref{Gauss_HD32}. The FWHM grows steadily with increasing wavelength as expected, from about $5$~mas around $8~\mu$m to $15$~mas for $13~\mu$m, depending on the baseline. These observations were performed at the orbital phases $0.19$, $0.2$ and $0.46$. This result suggests an elongated dust distribution.

\begin{figure}[h]
\vspace{-0.6cm}
\includegraphics[width=\hsize]{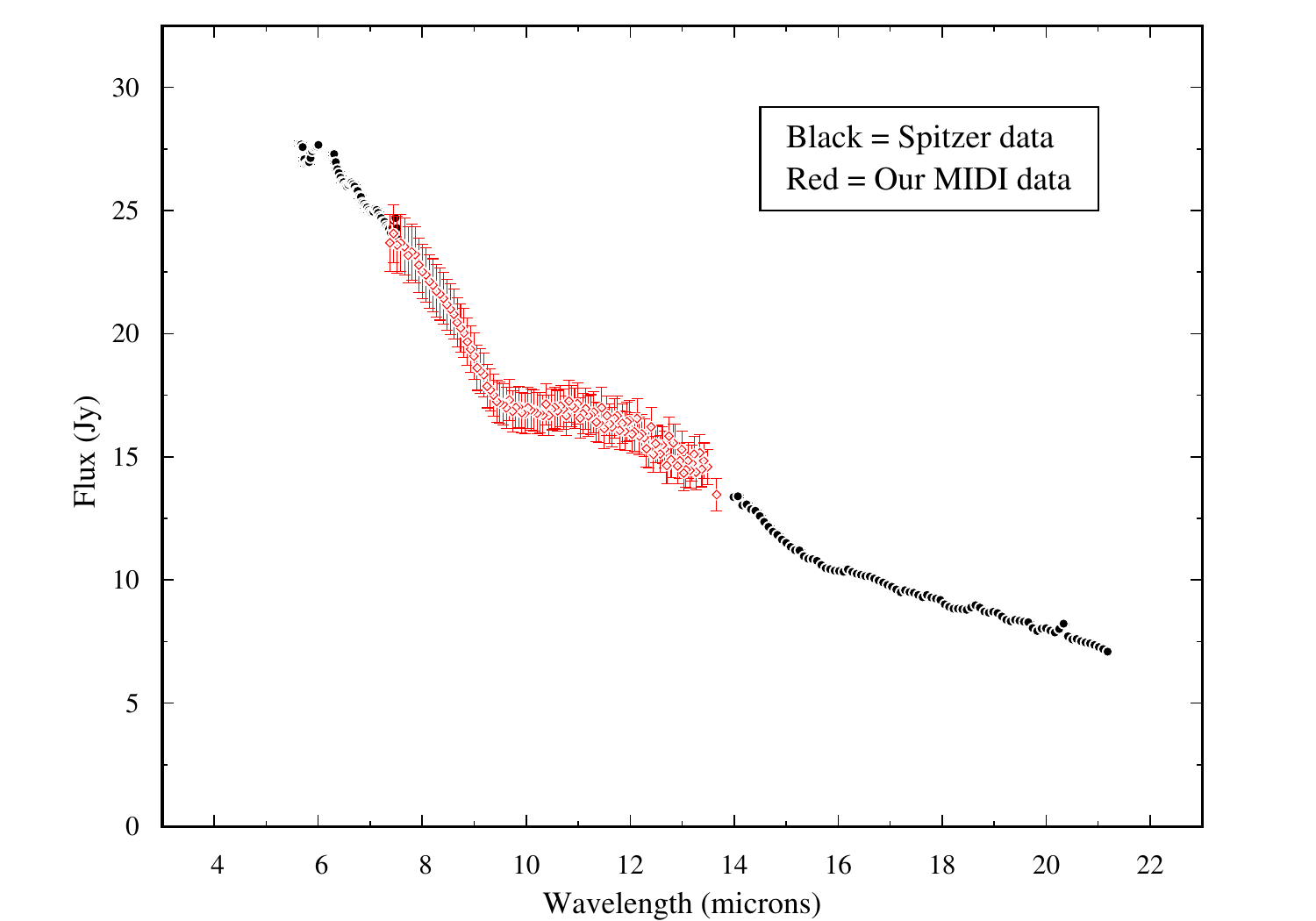}
\caption{Comparison between the MIDI spectral fluxes of \object{HD\,327083} (in red) with Spitzer archive data (in black). The amorphous silicate feature at $9.7\,\mu$m is seen in absorption.}
\vspace{-0.4cm}
\label{Spitzer_MIDI_HD32}
\end{figure}

Nevertheless, MIDI spectral fluxes reveal a conspicuous broad and deep absorption feature that peaks at $9.7~\mu$m attributed to stretching modes of Si-O bonds in amorphous silicates \citep{Roche1984}, shown with red symbols in Fig.~\ref{Spitzer_MIDI_HD32}. This figure displays the mean calibrated MIDI fluxes (with error bars between $10 \%$ and $20 \%$ of their corresponding fluxes) combined with  Spitzer data  (black symbols). The presence of this absorption feature is also confirmed by the absence of an increase of the FWHM around $10~\mu$m, as seen in Fig.~\ref{Gauss_HD32}.
  
\section{Discussion}
\label{Disc}
 To study the binary system and its evolutionary status, we carried out and analysed high-spectral and spatial resolution observations of \object{HD\,327083}, from optical to mid-IR ranges. We also derived the physical and chemical properties of the gaseous and dust discs surrounding the primary and the entire binary system.
 
 Furthermore, we improved the orbital parameters by combining our RV measurements with those obtained by \citet{Miroshnichenko2003}. We found an orbital period of $P=107.699$ days, already reported in \citet{Maravelias2018}, which provides a perfect phase-folded light curve when applied to the ASAS-3 observations.
 In addition, we found that the system has a very low eccentricity, indicating that tidal circularisation is almost reached. In parallel with our work, \citet{Porter2021} and \citet{Nodyarov2022,Nodyarov2024} determined periods of $107.991$ and $107.7$ days, respectively.
 We derived a mass function for the system of $1.357$, which is very similar to the $1.26$ value found by \citet{Nodyarov2024}.

\subsection{Properties of the binary system}

 We found that the binary system encompasses an accreting B-type star ($26\,000$~K, $\log\,g=3.0$) with a gaseous envelope and a cool giant of spectral type F6  ($6\,750$~K, $\log\,g=2.0$). A molecular disc and dust surround the binary system. The system is almost circularised. 
Due to the lack of photospheric 
absorption lines of the B-type star, we analysed its circumstellar gaseous disc that displays emission lines of \ion{Fe}{ii}. The RV curve based on these emission lines has almost the same amplitude as that of the late-type star ($K_2 \sim K_1$).
However, as the emission lines trace the dynamics of the star plus the disc, which is not necessarily circular,  the mass of the hidden component may be comparable to or even exceed that of the late-type star.

The light curve of \object{HD~327083} shows indications of tidal deformation, as described in Sect.~\ref{ASAS}. However, the spectrum of the cool companion agrees well with that of an F6 II–III star and appears typical for its spectral type. Therefore, the `deformation' of the star corresponds to a close filling of the high-mass component's Roche lobe and the onset of mass transfer \citep[][see also details in Sect.~\ref{nature}]{Vaidman2025}.
The excess material that cannot be accreted efficiently is likely expelled over the poles of the hot companion.
Such a scenario affects both the evolution of the binary orbit through the loss of mass and angular momentum, as well as the stellar components themselves.

To quantify the size of the Roche-lobe filling late-type star, we calculated the expected critical radius by applying Eggleton's relationship ~\citep{Eggleton1983}: 
\begin{equation}
  \frac{R_{L_1}}{a}= \frac{0.49 ~ q^{-\frac{2}{3}}}{0.6~q^{-\frac{2}{3}}+\ln(1+q^{-\frac{1}{3}})}, 
  \label{Eq:RL}
\end{equation}

\noindent 
where $a=a_1 +a_2$ is the distance between the two stars and $q = \frac{\mathfrak{M}_2}{\mathfrak{M}_1}$ is the mass fraction. 
To search for the best evolutionary model (see Sect.~\ref{nature}), we explored different values of $\mathfrak{M}_2$, ranging from those corresponding to inclination angles of $i=90 \degr$ \citep[seen edge-on,][]{Miroshnichenko2003} to $i=48.5 \degr$ \citep{Wheelwright2012b}. 
The latter was derived from AMBER data for the CO disc. Furthermore, assuming the donor star is filling the Roche lobe (Eq.~\ref{Eq:RL}), the stellar radii and bolometric magnitudes of each component are calculated. Table~\ref{Tab_new} lists possible values of $R_{L_{1}}$ (where $R_1=R_{L_{1}}$), $a$, and $M_{\rm bol}$, as well as the minimum and maximum masses of the stellar components, for a given $q$ and within the considered range of inclinations. We used Eq.~\ref{Eq:m} and the third Kepler's law.

\begin{table*}  
\caption{Binary system parameters for two orbital inclinations. \label{Tab_new}}
\begin{center}
\tabcolsep 3.0pt    
\begin{tabular}{c|cccccrc|rrcrcrc}
   \hline\hline
\noalign{\smallskip}
 &\multicolumn{7}{c|}{$i=90^\circ$} & \multicolumn{7}{c}{$i=48.5^\circ$}\\
 \noalign{\smallskip}
 \hline
 \noalign{\smallskip}
$q$ & $\mathfrak{M}_1$ & $\mathfrak{M}_2$ & $a$ & $R_1$ &  $M_{\rm bol~1}$ & $R_2$   & $M_{\rm bol~2}$ & $\mathfrak{M}_1$ & $\mathfrak{M}_2$ & $a$ & $R_1$ &  $M_{\rm bol~1}$ & $R_2$   & $M_{\rm bol~2}$\\
& [M$_\sun$] & [M$_\sun$] & [R$_\sun$] & [R$_\sun$] &  [mag] & [R$_\sun$] &  [mag]  & [M$_\sun$] & [M$_\sun$] & [R$_\sun$] & [R$_\sun$] &  [mag] & [R$_\sun$] &  [mag] \\
\noalign{\smallskip}
    \hline
    \noalign{\smallskip}
$0.9$	& 	$6.7$&	$6.0$&	 $223$&			$86.5$ &$-5.7$	& $ 12.1$	& $-7.5$ & $16.0$ &$14.4$&$297$&$115.5$ &$-6.3$&$16.1$&$-8.1$ \\
$1.0$	& $5.4$ &	$5.4$	& $211$ &			$80.0$ & $-5.5$ &	$11.2$ &	$-7.3$ &  $12.9$ &$12.9$&$281$&$106.0$& $-6.1$&$15.0$&$-7.9$
\\
$1.25$	& $3.5$&$4.4$ &$190$ &$68.3$ &$-5.2$ &$9.6$ & $-7.0$ & $8.4$ &$10.5$&$254$&$91.2$ &$-5.8$&$12.8$&$-7.6$\\	

$1.5$ &	$2.5$ &	$3.8$ & $176$ &$60.6$	&$-4.9$ &$8.5$ &$-6.7$ & $6.0$ &$9.0$&$235$&$80.9$ &$-5.5$&$11.3$&$-7.3$\\
$2.0$ &	$1.5$	& $3.0$ &	 $158$ &	$50.8$ &		$-4.5$ &	$7.1$	& $-6.3$ & $3.6$ &$7.3$&$211$&$67.8$ &$-5.2$&$9.5$&$-7.0$\\
\noalign{\smallskip}
\hline\hline
    \end{tabular}
    \tablefoot{The Table describes the stellar masses ($\mathfrak{M}_1$ and $\mathfrak{M}_2$), orbital separation ($a$), stellar radii ($R_1$ and $R_2$), and bolometric magnitudes ($M_{\rm bol~1}$ and $M_{\rm bol~2}$) of the binary components for two different orbital inclinations. The stellar radii represent the upper limit corresponding to the case in which the donor star fills its Roche lobe, $R_{L_{1}} = R_{1}$.}
    \end{center}
\end{table*}

\subsection{The circumbinary environment}

 The near and mid-IR spectrum reveals an O-rich environment. Both near-IR CO and SiO molecular features are in emission, while dust silicate features are in absorption. From fitting the CO emission, we confirmed that the molecular ring is formed by material ejected by the evolved star since it presents $^{13}$CO enrichment. The RVs of the CO band heads indicate that the molecular gas is circumbinary.

\citet{Wheelwright2012b} estimated a size of approximately $3 \pm 0.3$ AU (adopting a distance of 1.5 kpc) for the inner edge of this circumbinary ring, which is far
 beyond the orbits of the two components. 
However, using the updated Gaia distance of $2.448 \pm 0.145$~kpc, this location is revised, and a larger value of $4.89 \pm 1.73$~AU is obtained.

 We found that the CO ring is stable over the observing period of eleven years.
 However, contrary to \citet{Wheelwright2012b}'s results, we report here evidence for CO variability on short timescales (within one month, see Fig.~\ref{CO_phase} and \ref{fig:2010-2021}). The rotationally broadened ro-vibrational CO lines obtained with CRIRES ($\phi=0.78$) display profiles with slightly more pronounced red peaks than the blue ones. This situation appears to be the opposite one month later (at the orbital phase $\phi=0.12$) when the 2010 Phoenix spectrum (despite being a little bit noisier) shows a much stronger blue peak (shown in Fig.~\ref{CO_phase}, bottom panel, and in Fig.~\ref{fig:2010-2021} with orange line). The blue shoulder of the second band head arises from the superposition of many individual ro-vibrational lines, all having an asymmetric line profile with the blue peak more intense than the red one, which easily accounts for the difference in shape. As these phases coincide with the beginning and end of the broad minimum in the light curve, the late-type star might also hide parts of the CO ring.
 Moreover, at $\phi=0.81$, the 2017 Phoenix spectrum reveals a higher CO intensity relative to the continuum, which might be interpreted with a change in continuum contribution.

Furthermore, MIDI spectroscopy detected the presence of a deep silicate absorption feature. 
Stellar objects with deep silicate absorption features are rarely observed since silicate is primarily seen in emission from warm circumstellar dust surrounding oxygen-rich stars \citep{Mathis1990}. In addition, polycyclic aromatic hydrocarbon (PAH) features at $8.6$ and $11.3$~$\mu$m are not observed (see Fig.~\ref{Spitzer_MIDI_HD32}). 

The VLTI/MIDI visibility curves are flat, with values close to 0.8, and the observed FWHM values increase with wavelength. Such behaviour was also reported in the B[e] supergiant \object{CPD-57$^{\circ}$~2874} \citep{Domiciano2007}, which suggested that the dust continuum emission originates from an elongated structure. For \object{HD~327083}, the inner size of the dusty region extends from $12.5$~AU ($5$~mas for $8~\mu$m) to $29.4$~AU ($12$~mas for $13~\mu$m) at $\phi=0.19-0.20$ and from $22.5$~AU ($9$~mas for $8~\mu$m) to $44$ AU ($18$~mas for $13~\mu$m) at $\phi= 0.47$, assuming the Gaia distance. This result confirms that the dust structure is elongated (likely elliptical) and circumbinary and is therefore located in the CO ring's outer regions. The extent of the dust regions agrees with the hot dust shells used to fit the SED (see Fig.~\ref{Fig_SED}).

\subsection{The nature of HD\,327083}
\label{nature}

The temperature of the B-type object is high enough ($26\,000$~K) to ionise the environment embedding the star, which becomes detectable by emission lines. While the optical line spectrum and near-IR colours of \object{HD\,327083} suggest an early-type supergiant with emission lines, the hydrogen Pfund series is not seen in emission. This puts the star apart from all other B[e]SGs we had studied \citep[see, e.g.][]{Kraus2013, Oksala2013}, where so far usually Pfund and CO band emissions appear simultaneously. 

We calculated an absolute magnitude of the system to be $M_{\rm V} = -7.5$~mag, using the apparent magnitude $m_{V}=9.8$~mag, the Gaia distance ($d = 2.45$~kpc), and the obtained ISM $E(B-V) = 1.4$~mag and an extra absorption of $\sim 1$~mag due to the gaseous and dust shells. According to Eq.~(\ref{Eq:RL}), the  expected bolometric magnitude for an F-type star with $T_{\rm eff} = 6750$~K that fills its Roche lobe is between $M_{\rm bol} = -5.5$~mag  and $-6.1$~mag (see Table~\ref{Tab_new}, for $q\sim1$). The table lists two different inclination angles, while the magnitude range reflects the upper-limit cases. Then, the $M_{\rm bol}$ of the hot companion is between  $-7.3$~mag and $-7.9$~mag, assuming that its radius is $0.14\,R_1$ (where $R_1=R_{L_{1}}$). We used the code BSE \citep{Tout1997} that describes the evolution of a binary system, considering mass loss, mass transfer, and common-envelope evolution. We explored several combinations of $q$ values, using the solution of $\mathfrak{M}_1$ for different inclinations (from $i=90^\circ$ to $i=48.5^\circ$, see Eq.~\ref{Eq:q}). There are only a few models that fit the observed effective temperature, the expected luminosity, and the separation between the components. The most plausible evolutionary model corresponds to an interacting binary with similar initial masses, that is  $13.2$~M$_\sun$ and $12.9$~M$_\sun$. 
The system underwent an extended phase of mass transfer, and after approximately $15.4$~Myr, it evolved to the observed orbital configuration with current masses of $\mathfrak{M}_1=12.9$~M$_\sun$ and $\mathfrak{M}_2=12.7$~M$_\sun$. This solution agrees with an inclination angle of $48.5^{\circ}$ for the orbital plane relative to the observer's line of sight.
 However, to fulfil the condition of the surface gravity of $\log\,g = 2.0$ for the cool component, as was derived from plane-parallel models of stellar atmosphere, the Roche lobe should be filled only up to some $60-70\%$ compared to the case of a complete Roche lobe filling (see Table~\ref{Tab_new}, case $q=1$ and an inclination angle of $48.5^{\circ}$).
Whether the accretion disc of the hot star is fed by direct mass transfer from the cool star or via accretion from the circumbinary disc is not yet constrained.

Alternatively, using the Keplerian rotation velocity of the CO disc of $75\pm1$\,km\,s$^{-1}$ and its radius of $4.89\pm1.73$
AU, we estimate the central mass of the system of about $31\pm11$~M$_\sun$. This value also agrees with the total mass found from the evolutionary model of the binary system ($\sim 26$~M$_\sun$).

The BSE code also provides details about the
various evolutionary stages of our binary system.  At $15.42$~Myr, the F-type star expands, initiating a Roche-lobe overflow (RLOF). The mass transfer likely continues as the star further expands within its lobe. This stage is our actual scenario. Shortly after, the system will enter a blue straggler star phase, where the mass gainer rejuvenates and appears hotter and more massive than expected for its age, accompanied by a significant change in stellar mass. 
At approximately $15.43$~Myr, a common envelope phase would begin. During this stage, the system undergoes a rapid orbital tightening while ejecting a large amount of mass. By $16.4$~Myr, the hot companion will evolve and begin its own RLOF and common envelope phase. The system will likely end in a supernova explosion.

 An alternative model of evolution was computed by \citet{Vaidman2025} using the Mesa code, who derived a current system age of $13.6\pm0.1$~Myr. This solution leads to a massive binary system (with  $\mathfrak{M}_1=13$~M$_\sun$ and $11.50$~M$_\sun$).  The state of the system corresponds to a close filling of the high-massive component’s Roche lobe and the beginning of the mass transfer. After the mass transfer event, the mass of the post-primary drops to $5$~M$_\sun$, the post-secondary mass grows until $20$~M$_\sun$, and the binary will convert to a detached system with a long orbital period of $\sim 700$~days.

In the context of our evolutionary configuration, the formation of the circumbinary discs of CO and silicate dust takes place as a consequence of the mass transfer process and during the filling of the Roche lobe of the primary star (F-type) through the outer Lagrangian point L2 \citep[e.g.][]{Shu1979, Pejcha2016}. Such binaries are supposed to be circularised. 
Although the F-type star does not host CO in its atmosphere, molecules can form in the surrounding environment once the temperature drops below 5000 K. The observed isotopic CO ratio of only 20 can be achieved during the pre-red supergiant phase in rotating stars \citep{Kraus2025}.

\section{Conclusions}
\label{conclusions}

We have presented new multi-epoch high-resolution optical and near-infrared spectra of the putative B[e] supergiant \object{HD~327083} along with photometric data taken over a time of about ten years. The spectra are dominated by absorption lines from the late-type companion (F6~II-III) and pure emission lines as well as lines with P~Cygni profiles connected with a hot star. The emission lines can be interpreted as arising from a dense environment, most likely an accretion disc, in which the hot companion is embedded, and the P~Cygni profiles trace a bi-polar wind emanating from it.  This conclusion is based on the shape of the light curve, which further suggests that the late-type star is tidally deformed and most likely underfilling its Roche lobe. Furthermore, the circumbinary molecular ring \citep[also tilted at $48.5^{\circ}$,][]{Wheelwright2012a} is partially blocked by the F-type star at some orbital phases, leading to asymmetric line profiles. This circumbinary ring is most likely a consequence of mass transfer during the RLOF phase through the Lagrangian point L2.  Due to tidal variations, we expect the circumbinary disc to precess and the binary system orbit to have a non-zero eccentricity.
Unfortunately, we do not have good coverage of the CO emission over the complete orbit of the binary system in order to fully describe the temporal evolution of the circumbinary ring.

 The VLTI/MIDI visibility curves indicate that the dust continuum emission originates from an elongated circumbinary structure located in the CO ring's outer regions. In addition, our MIDI spectrum reveals a strong absorption band of amorphous silicates, which is often seen in oxygen-rich stars. Follow-up interferometric observations over the orbital period are clearly needed to confirm the observed variable distance of the inner disc rim with the orbital phase and to unveil its nature.

Our analysis of the radial velocities of numerous lines resulted in a refined orbital solution of the B[e] binary system. We derived a period of $P = 107.699$~d, in agreement with \citet{Nodyarov2024}'s results.
In addition, the emission lines exhibit variations in both shape and intensity throughout the orbital period, likely caused by attenuation from the gas stream.
Furthermore, we found that both stars have similar masses of $12.9$~M$_\sun$, and the system has a separation of $1.3$~AU.

From our analysis, we conclude that the B[e] binary system \object{HD~327083} is nearly circularised and would be made up of stars of similar masses (a supergiant of $26\,000$~K, early B type) and a Roche lobe partially filled by the F-type companion that is probably evolving towards the red supergiant stage.

\begin{acknowledgements}
We would like to thank the referee for carefully reviewing our manuscript and providing thoughtful and constructive comments. This work is based 
on observations taken with 1) Telescopes at Paranal ESO Observatory under the program  $085.D-0454$ for CRIRES,  and $085.D-0454(B)$ for MIDI; 2) The MPG/ESO 2.2-m telescope and the ESO 1.52-m telescope under programs $385.D-0513(A)$, $075.D-0177(A)$, $082.A-9209(A)$, $085.D-0185(A)$, $094.A-9029(D)$, $095.A-9032(A)$, $096.A-9024(A)$, $096.A-9030(A)$, $096.A-9039(A)$, $097.A-9024(A)$ and $097.A-9039(C)$ for FEROS; 3) Gemini South/Phoenix instrument, science program  $GS-2010A-Q-41$, $GS-2017A-Q-30$ and $GS-2021A-Q-401$; 4) J. Sahade Telescope at Complejo Astron\'omico El Leoncito (CASLEO), operated under an agreement between the Consejo Nacional de Investigaciones Cient\'{\i}ficas y T\'ecnicas de la Rep\'ublica Argentina, the Secretar\'{\i}a de Ciencia y Tecnolog\'{\i}a de la Naci\'on and the National Universities of La Plata, C\'ordoba and San Juan; 5) The Observat\'orio do Pico dos Dias, LNA, Brazil. The work also used the Immersion Grating Infrared Spectrometer (IGRINS) that was developed under a collaboration between the University of Texas at Austin and the Korea Astronomy and Space Science Institute (KASI) with the financial support of the US National Science Foundation 27 under grants AST-1229522 and AST-1702267, of the University of Texas at Austin, and of the Korean GMT Project of KASI.
The FEROS observations obtained between 2014 and 2017 with  the MPG 2.2m telescope were supported by
the Ministry of Education, Youth and Sports project - LG14013 (Tycho Brahe: Supporting Ground-based Astronomical Observations). We want to thank the observers (S. Ehlerova, A. Kawka, and L. Zychova) for obtaining the data. In addition, 
this research was achieved using the POLLUX database (http://pollux.oreme.org/), operated at
LUPM (Université de Montpellier - CNRS, France) with the support of the PNPS and INSU. This work is also based in part on observations made with the Spitzer Space Telescope, which was operated by the Jet Propulsion Laboratory, California Institute of Technology under a contract with NASA.
The Astronomical Institute in Ond\v{r}ejov is supported by the project RVO:67985815.
MLA and AFT acknowledge financial support from CONICET (PIP 1337) and the Universidad Nacional de La Plata (Programa de Incentivos 11/G160), Argentina. YJA thanks funding from the Universidad Nacional de La Plata (Programa de Incentivos 11/G162), Argentina. MBF acknowledges financial support from the National Council for Scientific and Technological
Development – CNPq – Brazil (grant number: 307711/2022-6). MC acknowledges the support from Centro de Astrofísica de Valparaíso, and MC and IA thank the support from the FONDECYT project 1230131.
This project has received funding from the European Union's Framework Programme for Research and Innovation Horizon 2020 (2014-2020) under the Marie Sk\l{}odowska-Curie grant agreement No. 823734 (POEMS) and HORIZON TMA MSCA Staff Exchanges grant agreement No. 101183150 (OCEANS).

\end{acknowledgements}

\bibliographystyle{aa}
\bibliography{cites}

\end{document}